\documentclass[a4paper,12pt]{article}
\usepackage[english]{babel}

\usepackage[T1]{fontenc}
\usepackage[top=2.6cm,left=2cm,right=2cm,bottom=2.5cm]{geometry}

\usepackage{ragged2e}
\usepackage{lmodern}
\usepackage{multicol}
\usepackage{amsmath}
\usepackage{amsthm}
\usepackage{amsfonts}
\usepackage{thmtools}
\usepackage{mathtools}
\usepackage{amssymb}
\usepackage{bm}
\usepackage{dsfont}
\usepackage{enumitem}
\usepackage{graphicx}
\usepackage[dvipsnames,table]{xcolor}
\usepackage[bottom]{footmisc}
\usepackage{hyperref}
\usepackage[nodisplayskipstretch]{setspace}
\usepackage{subcaption}
\usepackage{fancyhdr}
\usepackage[normalem]{ulem}
\usepackage{tabularx}
\usepackage{booktabs}
\usepackage{multirow}
\usepackage{makecell}
\usepackage{xurl}
\usepackage{wasysym}

\usepackage[authoryear,round,comma]{natbib}

\allowdisplaybreaks

\newcommand{\E}{\mathds{E}}

\newcommand{\Int}{\mathbb{Z}}
\newcommand{\Real}{\mathbb{R}}

\newcommand{\1}[1]{\mathbb{I}\{#1\}}

\newcommand{\norm}[1]{\left\lVert{#1}\right\rVert}
\renewcommand{\d}{\textnormal{d}}

\theoremstyle{plain}
\newtheorem{theorem}{Theorem}[section]

\theoremstyle{definition}

\theoremstyle{remark}
\newtheorem{remark}[theorem]{Remark}

\definecolor{blue}{HTML}{0072BD}
\definecolor{green}{HTML}{77AC30}
\definecolor{orange}{HTML}{D95319}
\definecolor{red}{HTML}{A2142F}
\definecolor{yellow}{HTML}{EDB120}
\definecolor{cyan}{HTML}{4DBEEE}
\definecolor{purple}{HTML}{7E2F8E}

\hypersetup{
	colorlinks=true,
	linkcolor=blue,
	citecolor=blue,
	urlcolor=blue
}

\begin{document}

\title{\normalfont%
    Multi-Horizon Echo State Network Prediction of\\
    Intraday Stock Returns
}
\author{%
Giovanni Ballarin$^{1}$, Jacopo Capra$^{2}$, and Petros Dellaportas$^{3}$
}
\maketitle

\renewcommand{\abstractname}{\vspace{-\baselineskip}} %
\begin{abstract}
	Stock return prediction is a problem that has received much attention in the finance literature. In recent years, sophisticated machine learning procedures have been shown to perform significantly better than conventional prediction techniques. One downside of these approaches is that they are often numerically expensive to implement, for both training and inference, because of their high complexity. We propose a return prediction framework for intraday returns at multiple horizons based on Echo State Network (ESN) models, wherein a large subset of parameters is drawn at random and never trained.
    The coefficients that do require training can be fitted by explicitly solving a rolling-window, ridge-regularized least squares problem.
    Our approach enjoys the flexibility of machine learning methods, has an inherently efficient implementation, and shows strong forecasting performance.
\end{abstract}
\bigskip
 

\makeatletter
\addtocounter{footnote}{1} \footnotetext{%
    Universit\"at Sankt Gallen, Mathematics and Statistics Division, Switzerland -- {\texttt{giovanni.ballarin@unisg.ch}}.
    GB is grateful for the support of the Center for Doctoral Studies in Economics and the Chair of Empirical Economics at the University of Mannheim, where part of this work was developed, as well as the Great Minds Postdoctoral Fellowship program at Universit\"at Sankt Gallen.
    Many thanks go to Lyudmila Grigoryeva and Erik-Jan Senn for their helpful comments and suggestions.
}
\addtocounter{footnote}{1} \footnotetext{%
    UCL, Department of Statistical Science, United Kingdom -- {\texttt{jacopo.t.capra@gmail.com} }}
\addtocounter{footnote}{1} \footnotetext{%
    UCL, Department of Statistical Science, United Kingdom and Athens University of Economics and Business, Department of Statistics, Greece -- {\texttt{p.dellaportas@ucl.ac.uk}}}
\makeatother
\maketitle
\setcounter{footnote}{0}%
\newpage

\section{Introduction}

The prediction of time series is a topic of great relevance in many economic and financial settings.
Researchers aim to devise more effective, reliable, and parsimonious forecasting methods and models. 
In the field of financial portfolio management, the prediction of stock features (such as price, return, volatility, and trading volume) is a key component of many trading strategies. 
Traditionally, linear models are used for this purpose, as they are straightforward to implement and analyze. 
Major developments in the fields of artificial intelligence, machine learning (ML), and statistical learning methods have, however, greatly expanded the universe of potential forecasting approaches. 
A variety of methodologies have been proposed to improve the prediction of stock market data \citep{kumbure2022machine}. Efforts involving neural networks (NNs) date back to the 1990s; see e.g. \cite{refenes1994stock,ghosn1996multi,saad1998comparative}, and \cite{zhangForecastingArtificialNeural1998}. 
In recent years, researchers have evaluated financial machine learning forecasting approaches from both empirical \citep{sirignanoUniversalFeaturesPrice2019,aletiIntradayMarketReturn2025} and theoretical perspectives. 
From the latter point of view, works such as \cite{didisheimComplexityFactorPricing2023} and \cite{kellyVirtueComplexityReturn2024} have formally argued why and how more complex models can much improve prediction, the so-called ``virtue of complexity'' \citep{kellyVirtueComplexityReturn2024}.
Another aspect to consider is that returns at different horizons may embed distinct predictive information, which an investor can exploit, e.g., by constructing portfolios jointly targeting asset returns at multiple horizons.
Multi-period portfolio optimization is thus a potentially attractive extension of the seminal mean-variance optimization (MVO) model developed by \cite{markowitz1952portfolio}. 
The classical MVO approach is based on a single-period model, and multi-period optimization can be seen as a generalization of the standard Markowitz setting; 
see \cite{blake2003theRightTool,grinold2007dynamic,qian2007informationHorizon,grinold2010signal,sneddon2008theTortoise,garleanu2013dynamic}.

In this work, we develop a novel, computationally efficient forecasting approach for intra-day, multi-horizon return prediction that is scalable to both (very) high-frequency data, as well as data-poor settings.\footnote{In our context, a ``\textit{data-poor}'' setting is one where, although potentially large volumes of data are available, only a small fraction of it can actually be used to fit a prediction model. This may be due, for example, to fast regime changes, missing data, or numerically expensive preprocessing steps.}
Given the effectiveness and flexibility of machine learning methods in capturing nonlinear relationships between data, and their success and recent widespread adoption in computer science, statistics, and econometrics, we wish to apply ML techniques to the intra-day multi-horizon return prediction problem. Yet, models such as deep neural networks are generally both expensive to train and require large datasets.
Instead, we consider a special class of recurrent-type neural networks (RNNs) that are only partially trained: The recurrent state parameters are kept fixed at their random initialization values. We leverage the so-called Reservoir Computing approach and apply \textit{Echo State Network} (ESN) models \citep{jaeger2001echo,lukoseviciusReservoirComputingApproaches2009}. 
The choice of applying the Echo State Networks to an intraday systematic trading context is also motivated by the empirical success ESNs have had in prediction tasks attempted in other fields \citep{tanakaRecentAdvancesPhysical2019}.
Our methodology builds on the recent advances in using ESNs for macroeconomic forecasting at different horizons and frequencies \citep{Ballarin2024reservoirComputingMacro}.  
When using a linear readout layer, ESNs can be fitted to data via regularized linear regression, which has a closed-form solution, unlike the computationally intensive stochastic backpropagation procedures used for neural networks. 
Three properties of ESN models are particularly useful in our setting: the \textit{expressivity} of Echo State Network models, which are universal in terms of approximating nonlinear time series processes; ESN adaptability to \emph{multi-horizon forecasting} by virtue of a nonlinear state formulation; the \emph{computational efficiency} of ESNs, which lets one solve high-dimensional data problems with comparatively negligible computational overhead. 

Our proposed ESN approach is predicated on first constructing a recurrent state equation that nonlinearly combines signals over time. The target returns are then predicted with a linear function of the ESN states, again using ridge-regularized least squares. 
For our regularized models, we employ a within-day rolling window cross-validation strategy to select hyperparameters. This setup allows us to systematically evaluate the performance gains of including regularization in the linear model, and additionally those obtained as a result of the ESN model. 

In our application, we target 1290 U.S. stock price returns across 5 different intraday horizons at 10-minute intervals within the trading session. 
Our full forecasting pipeline can be run over one calendar year of data in a handful of minutes, highlighting the numerical efficiency of our ESN proposal.
To measure forecasting effectiveness, we consider both mean squared forecasting error (MSFE) and the $R^2$ of predicted returns. 
As benchmarks, we employ two linear forecasting models, as well as a machine learning model. 
The linear baselines are based on, respectively, ordinary linear regression over a vector of signals based on past market and stock performance, and a ridge-regularized linear model that shrinks coefficient estimates over the same set of signals. 
As ML benchmark, we also include a random forest model that takes the signals as inputs \citep{breimanRandomForests2001}.
Our results show that our ESN-based forecasting strategy for the cross-section of returns dominates the benchmarks at all intraday horizons, with only the case of end-of-day (close) returns showing rather modest improvements. 
Notably, we achieve an up to 0.87\% MSFE reduction that is also remarkably robust to the inherent model sampling randomness of ESN models. The random forest's performance is, however, \textit{worse} than that of the linear baseline by approximately the same absolute margin as the ESN.
Per-horizon Diebold-Mariano and model confidence set (MCS) tests confirm that these gains are all statistically significant. 
Finally, to gauge the economic utility of our forecasts, we construct MV portfolios for a 10-minute horizon, non-overlapping returns, including a benchmark equal-weights portfolio. We find that ESN predictions yield the best gross cumulative returns and Sharpe ratio, and the lowest transaction costs and drawdowns.

\paragraph{Contributions.}
The core contributions of our work can be summarized as follows
\begin{itemize}
    \item We provide a highly efficient, robust, and effective machine learning prediction framework for high-frequency returns that can be applied to very large asset panels. ESN model estimation and hyperparameter validation runtimes are comparable to those of (regularized) linear regression, enabling large-scale real-time model optimization.
    \item We rigorously test the performance of our proposed model against benchmarks, including random forests, which have been extensively used in the prediction of tabular data. 
    Our empirical results show that ESNs achieve better forecasts than RFs at a much lower computation cost, and these prediction improvements also translate to higher returns and lower turnover in the respective MV strategies. 
    \item Our empirical results provide additional insights on how predictability changes at different intraday horizons for a large cross-section of U.S. stocks. They suggest that this phenomenon cannot be readily circumvented by relying on nonlinear prediction schemes, as their MSFEs eventually concentrate close to those of linear prediction models.
    \item Since, as we explain in detail in Section~\ref{section:esn_models}, ESN models have an inherently random nature, it is not known a priori how stable the forecasting performance is. While this aspect is often overlooked in practical applications of ESNs, we re-evaluate our empirical results for many different draws of the models. Our analysis shows that the prediction performance of our proposed approach is robust to model randomness.
\end{itemize}

\paragraph{Outline.}

In Section~\ref{section:literature}, we provide a review of the recent literature on ML methods in finance.
Precise descriptions of the data, trading setting, and benchmark models are given in Section~\ref{section:data_implement}. 
Section~\ref{section:esn_models} presents our proposed ESN model to forecast stock returns based on user-provided signals (we shall also use the word ``features'' interchangeably), as well as training procedures.
Section~\ref{section:application}, where we discuss our application to U.S. stock returns prediction over the year 2013, follows.
Finally, Section~\ref{section:conclusion} concludes.

\section{Related Literature}\label{section:literature}

Classical modeling of financial markets based on efficiency and risk neutrality suggests that stock returns should be unpredictable, and as such they are often econometrically modeled as random walks \citep{pesaranTimeSeriesPanel2015}. Empirical finance research has nonetheless shown that this may not be entirely the case, with some evidence pointing towards the potential violation of these assumptions \citep{campbellEconometricsFinancialMarkets2012}. 
Focusing on intra-day returns specifically, \cite{ait-sahaliaHowWhenAre2025} recently provided extensive evidence of predictability at millisecond scales. 
Moreover, \cite{aletiIntradayMarketReturn2025} showed that intraday returns may be effectively forecasted through the construction of a large collection of factors.
We tackle a similar setting, with data sampled at 10-minute intervals, and find that at this coarser scale, too, there appears to be non-negligible return predictability.\footnote{In terms of \textit{volatility} forecasting, \cite{dhaene2020incorporating} show that intraday (5-minute resolution) and overnight returns can also be incorporated in multivariate GARCH and BEKK-type models to considerably sharpen prediction.} 
It should be underscored that, compared to the earlier works mentioned below, to our knowledge, ours is the first to (i) consider a large universe of stocks, larger than e.g. the S\&P500 components, that forms an \textit{unbalanced} panel with missing observations, (ii) work at high (intra-day) frequencies, (iii) target multiple overlapping return horizons and (iv) implement a rolling-window cross-validation and fitting algorithm to allow for fast data regime changes.

\paragraph{ML Methods in Finance.}

There are many works in the past two decades that implement various machine learning techniques in a financial setting.
\cite{olson2003neural} and \cite{kwon2007aHybrid} both employ modern neural network approaches to stock forecasting.
\cite{yu2020stock} more recently employed {Deep} Neural Networks (DNNs) for the same task, finding that the flexibility afforded by deep models improves performance.
\cite{borovkova2019anEnsembleLSTM} use a Long Short-Term Memory (LSTM) recurrent architecture to classify stocks in high-frequency.
\cite{tolo2020predicting} exploits RNN-type networks to predict systemic crises in the financial sector.
With a view towards the structural changes due to the COVID-19 pandemic, \cite{chandra2021bayesian} resort to a Bayesian NN approach to forecast prices of stocks from exchanges in the United States, Germany, China, and Australia, while \cite{abedin2021deep} use a more standard deep learning approach to perform exchange rate prediction. 
Further, \cite{wang2024factorGAN} propose a factor-based Generative Adversarial Network (GAN) to tackle both prediction and portfolio investment problems.
For MVO specifically, \cite{kim2024enhancing} also applied GANs: Their idea is to perform anomaly detection and augment the portfolio optimization problem with said information. 
\cite{vuleticFinGANForecastingClassifying2024}, using a novel ``Fin-GAN'' model, have shown that GANs trained with an economics-driven loss can be effective in achieving higher Sharpe ratios compared to both LSTM and ARIMA models.
Lastly, it is important to mention that ML models, such as GANs and Variational Autoencoders, have also been extensively studied for synthetic financial data generation, see e.g. \cite{kim2023aGANs-Based,masi2023onCorrelated,nagy2023generative,acciaio2024time,kwon2024can}, and \cite{cetingoz2025synthetic} for a theoretical critique of this line of research.

\paragraph{Reservoir Computing Applications.}

Sophisticated neural network models are challenging to implement and train by construction. 
Echo State Networks are a recent development in the field of machine learning models that aim to tackle this issue, while still matching the flexibility of fully-trainable neural networks.
By avoiding training of state parameters, ESNs and RC models are able to circumvent the issues associated with backpropagation-through-time associated with classical RNNs \citep{hochreiterVanishingGradientProblem1998,vlachasBackpropagationAlgorithmsReservoir2020}.
Echo state networks have been applied to a wide variety of contexts \citep{sunSystematicReviewEcho2024}. 
For example, ESNs have been successfully employed to predict water levels \citep{coulibaly2010reservoirGreatLakes}, electricity loads \citep{deihimi2012echoStateElectricityLoad}, and power generation by renewable sources \citep{hu2020forecastingEnergyConsumptionDeepESN}. Variations involving deep architectures have also been devised \citep{kim2020deepESNTimeSeries}.
In the context of economic forecasting, \cite{Ballarin2024reservoirComputingMacro} constructed a class of ESN models that can incorporate data sampled at different frequencies, outperforming state-of-the-art models when forecasting U.S. GDP growth.
The literature on financial applications is, to the best of our knowledge, minimal. 
\cite{linShorttermStockPrice2009,linIntelligentStockTrading2011} were among the first to apply ESNs to stock price forecasting at daily frequency. Their model specification included a noise component in the state (which is known to negatively affect memory of past inputs, c.f. \citealp{guanHowNoiseAffects2025}), and used a fixed train-test split.
Works such as \cite{zhangStockPredictionBased2013} and \cite{tianStocksPricePrediction2024} explored different avenues of extension of the ESN approach---phase space transformations and swarm-type hyperparameter search, respectively---but always in small-scale empirical setups.\footnote{\cite{zhangStockPredictionBased2013} evaluate their approach only on a single stock (Microsoft Corporation), while \cite{tianStocksPricePrediction2024} consider 10 U.S. and 10 Chinese stocks. Both papers use data collected at a daily frequency.}
Recently, \cite{sharmaEchoStateNetworks2025} showcased the efficacy of ESNs in predicting daily Bitcoin closing prices. 
\cite{liu2018ESNFinancialDataForecasting} provide an ESN application to financial data, but primarily in the context of hyperparameter optimization, while \cite{trierweilerribeiroNovelHybridModel2021} use ESN models for the prediction of stock return volatilities. \citep{macielExchangeRateForecasting2014} used Echo State Networks on currency exchange rate forecasting and trading.
Lastly, alternative models also based on randomization have also been considered; see e.g. \cite{akyildirim2023randomized} for \textit{random signatures} applied to portfolios and \cite{gonon2023random} for theoretical guarantees on \textit{random shallow network} learning of Black-Scholes PDEs.
To the best of our knowledge, we are the first to quantitatively analyze the application of ESNs to forecasting and portfolio construction problems for a large panel of stocks.

\section{Data and Setup}\label{section:data_implement}

Intraday stock data was sourced from AlgoSeek, a market-leading platform providing high-quality high-frequency trading data.\footnote{\url{https://www.algoseek.com/}} 
The data provider provides access to up to 1-minute resolution OHLC (open, high, low, close price) bars built from the reported Trades of the SIP (consolidated) data feed, which includes all U.S. Exchanges and FINRA from January 2007 to October 2020.
AlgoSeek bars cover the entire trading day:
\begin{itemize}
    \item Pre-Market: 4:00:00~AM to 9:29:59~AM;
    \item Market: 09:30:00~AM to 4:00:00~PM;
    \item Extended hours: 4:00:01~PM to 8:00:00~PM.
\end{itemize}

Throughout our analysis, we focus only on prices observed during market hours and employ Eastern Standard Time (EST) data time-stamps, although actual time and date may be different for each exchange due to geographical location. 
Given the large volume of data associated with the 1-minute frequency, we use downsampled observations at a 10-minute resolution in our empirical exercises.
Nonetheless, our forecasting approach can be readily scaled to higher-frequency data.
For an extended description of the data and its preprocessing, we refer the reader to Appendix~\ref{appendix:data}.

\subsection{Trading Environment}

The design of our forecasting experiments aims to mimic the setup of an intraday trading book, which rebalances its position only on an intraday basis and does not keep open the positions overnight. Accordingly, we assume that a portfolio rebalance can take place every 10 minutes, from 9:30~AM to 3:50~PM. 
Until 3:50~PM, we run individual stock return predictions every 10 minutes. It is important to note that all our predictions are done over future \emph{overlapping} return horizons. For example, at 9:30~AM, the 10-minute prediction horizon goes from the 9:31~AM price bar to the 9:41~AM price bar. On the other hand, the 30-minute prediction horizon goes from the 9:31~AM price bar to the 10.01~AM price bar. 

We target the following prediction horizons:
\begin{itemize}
    \item \emph{10 minutes} \textnormal{(10 min)}: 39 predictions per day over this horizon.\footnote{Returns are constructed at 10-minute intervals from 9:30 AM to 3:50 PM, which is the last time-stamp available for an intraday 10-minute horizon prediction. At 3:50 PM, the targeted return takes the closing auction price into account for the return calculation. The number of possible return predictions for other horizons is similarly computed, apart from end-of-day returns. For the \textsc{eod} horizon, returns are still constructed every 10 minutes, but they are overlapping and cannot be observed within the trading day.}
    \item \emph{30 minutes} \textnormal{(30 min)}: 37 predictions per day over this horizon.%
    \item \emph{60 minutes} \textnormal{(60 min)}: 34 predictions per day.%
    \item \emph{2 hours} \textnormal{(2 hours)}: 28 predictions per day.%
    \item \emph{End-of-day} \textnormal{(\textsc{eod})}: 39 predictions per day. Specifically, \textsc{eod} returns are built from the price at a particular time-stamp and the daily closing price. %
\end{itemize}

We do not carry out any return prediction at 4:00~PM, but rather assume that we fully exit our book positions, submitting our orders at 3:50~PM to be executed at the closing auction price. Accordingly, we do not target any continuous session close to auction and overnight return horizons.

\subsection{Prediction Setup}\label{subsection:data_setup}

The universe of tradeable stocks is indexed by $i = 1, \ldots, N$. 
For each stock, we have access to both future returns, $r^{(i)}_{t+h}$, and a set of $D$ signals, collected in a $D$-dimensional vector $Z^{(i)}_{t} = (Z^{(i)}_{1,t}, \ldots, Z^{(i)}_{D,t} )' \in \Real^{D}$. We shall also assume that $D$ is fixed over time.
Note that $r^{(i)}_{t+h}$ is indexed by horizon $h$, which for our purposes is an arbitrary intraday interval of time.
For example, at any given time $t$, we write $h = \textnormal{30 min}$ to indicate the 30-minute-ahead return, and $h = \textnormal{\textsc{eod}}$ for the end-of-day return.
If the $i$-th stock return or any of its signals are not available at time $t$, we discard it for period $t$ in our empirical exercise.\footnote{Numerically, if a stock drops from the panel at a certain date or data is missing, we use \texttt{NaN}s to represent its signals and/or returns for that time index. 
} 
Therefore, we assume that the \textit{observable} cross-section at time $t$ is composed of stocks with indices in set $\mathcal{I}_t \subseteq \{1, \ldots, N\}$ of size $N_t := |\mathcal{I}_t|$, $N_t \leq N$.

Our per-horizon forecasting objective is the prediction of the conditional expected return 
\begin{equation*}
    \E\big[ r^{(i)}_{t+h} \,\big\vert\, \mathcal{F}^{(i)}_t \big] ,
    \quad\text{where}\quad
    \mathcal{F}^{(i)}_t := \sigma(Z^{(i)}_t, Z^{(i)}_{t-1}, \ldots)
\end{equation*}
and $\sigma(Z^{(i)}_t, Z^{(i)}_{t-1}, \ldots)$ is the sigma algebra generated by all past realizations of $Z^{(i)}_t$, that is, the history of stock-specific signals. We are going to assume throughout that $r^{(i)}_{t+h}$ is measurable with respect to this filtration, and that any observable predictive information associated with past returns is already fully embedded in the signals.

\subsection{Signal Construction}
\label{subsection:signals}

Our primary goal in signal construction is to separate stock-specific features from the general market structure. Moreover, while we could, in principle, work with discrete signals, such as buy-to-open or sell-to-open, and combinations of these with continuous signals. 
For the sake of simplicity, we focus on the systematic construction of real-valued signals following \cite{avellanedaStatisticalArbitrageUS2010}. We discuss in more detail the technical aspects of the construction of signals in Appendix~\ref{section:signal_construction}. 

Let $P^{(i)}_t$ be the individual asset price. First, assume that prices are determined as a continuous-time stochastic process. Returns can be decomposed into drift, market factor (systematic), and idiosyncratic components:
\begin{equation}\label{eq:factor_model}
    \frac{\d P^{(i)}_t}{P^{(i)}_t}
    =
    a^{(i)} \d t + \sum_{j=1}^J b^{(i)}_{j} F_{j,t} + \d U^{(i)}_t ,
\end{equation}
where $a^{(i)}$ is the stock price drift, $\{ F_{j,t} \}_{j=1}^J$ are the returns of $J$ risk-factors associated to the market, $\{ b^{(i)}_{j} \}_{j=1}^J$ are the $J$ factor loadings and $\d U^{(i)}_t$ is a stock-specific residual component. 
We then model the residual term as an Ornstein-Uhlenbeck (OU) process:
\begin{equation}\label{eq:OU_process}
    \d U^{(i)}_t 
    = 
    \kappa^{(i)} \big( m^{(i)} - U^{(i)}_{t} \big) \d t + \sigma^{(i)} \d W_{t} ,
\end{equation}
where $W_{t}$ is the standard Wiener process.
As in \cite{avellanedaStatisticalArbitrageUS2010}, $\kappa^{(i)}$, $m^{(i)}$, and $\sigma^{(i)}$ are stock-specific parameters which are only slowly varying over time. 

The return residual model given by
\eqref{eq:factor_model}--\eqref{eq:OU_process} is implemented in
discrete time using a two-step procedure. First, we extract the market
factors from observed returns using principal component analysis (PCA).
For each stock $i$, we then estimate the factor regression
\begin{equation}
    r^{(i)}_t
    =
    a^{(i)}
    + \sum_{j=1}^J b^{(i)}_j F_{j,t}
    + \upsilon^{(i)}_t ,
    \label{eq:discrete_factor_model}
\end{equation}
and denote the corresponding estimated idiosyncratic return residuals by
$\widehat{\upsilon}^{(i)}_t$.
For a chosen lookback window $P$, we construct the cumulative residual
process
\begin{equation*}
    \widehat{U}^{(i)}_{P,t}
    :=
    \sum_{s=t-P+1}^{t} \widehat{\upsilon}^{(i)}_s .
    \label{eq:cumulative_residual}
\end{equation*}
The continuous-time OU specification in \eqref{eq:OU_process} implies
that, when observations are one sampling interval apart, its exact
discrete-time representation is an AR(1) process. We therefore estimate
\begin{equation}
    \widehat{U}^{(i)}_{P,t+1}
    =
    c^{(i)}_{0,P}
    +
    c^{(i)}_{1,P}\widehat{U}^{(i)}_{P,t}
    +
    \eta^{(i)}_{P,t+1}.
    \label{eq:residual_ar1}
\end{equation}
Provided that $0<c^{(i)}_{1,P}<1$, the corresponding estimates of the
mean-reversion rate and long-run mean are
\begin{equation*}
    \widehat{\kappa}^{(i)}_P
    =
    -\log c^{(i)}_{1,P}
    \qquad\textnormal{and}\qquad
    \widehat{m}^{(i)}_P
    =
    \frac{c^{(i)}_{0,P}}
         {1-c^{(i)}_{1,P}} .
    \label{eq:ou_parameters}
\end{equation*}
Let
\begin{equation*}
    \widehat{q}^{(i)}_P
    :=
    \widehat{\mathrm{Var}}\!\left(\eta^{(i)}_{P,t}\right).
\end{equation*}
denote the estimated innovation variance in
\eqref{eq:residual_ar1}. The stationary standard deviation of the fitted
discrete-time process is then
\begin{equation*}
    \widehat{\sigma}^{(i)}_{U,P}
    =
    \sqrt{
        \frac{\widehat{q}^{(i)}_P}
             {1-\left(c^{(i)}_{1,P}\right)^2}
    } .
    \label{eq:ou_stationary_sd}
\end{equation*}
Consequently, the standardized residual, or \emph{$z$-score signal},
associated with lookback window $P$ is
\begin{equation}
\label{eq:z_score_signal}
    z^{(i)}_{P,t}
    :=
    \frac{
        \widehat{U}^{(i)}_{P,t}
        -
        \widehat{m}^{(i)}_P
    }{
        \widehat{\sigma}^{(i)}_{U,P}
    } .
\end{equation}
This signal measures the deviation of the stock-specific cumulative
residual from its estimated long-run mean in units of its stationary
standard deviation.

The factor model in \eqref{eq:discrete_factor_model} also contains the
stock-specific drift $a^{(i)}$. To incorporate this component, note that
the conditional drift of the idiosyncratic component in the
continuous-time model is
\begin{equation*}
    a^{(i)}
    +
    \kappa^{(i)}
    \bigl(m^{(i)}-U^{(i)}_t\bigr)
    =
    -\kappa^{(i)}\sigma^{(i)}_U
    \left[
        \frac{U^{(i)}_t-m^{(i)}}{\sigma^{(i)}_U}
        -
        \frac{a^{(i)}}{\kappa^{(i)}\sigma^{(i)}_U}
    \right].
\end{equation*}
This motivates the \emph{modified $z$-score signal}
\begin{equation}
\label{eq:mod-z-score_signal}
    \widetilde{z}^{(i)}_{P,t}
    :=
    z^{(i)}_{P,t}
    -
    \frac{
        \widehat{a}^{(i)}
    }{
        \widehat{\kappa}^{(i)}_P
        \widehat{\sigma}^{(i)}_{U,P}
    } ,
\end{equation}
where $z^{(i)}_{P,t}$ is as in \eqref{eq:z_score_signal} and  $\widehat{a}^{(i)}$ is obtained from
\eqref{eq:discrete_factor_model}. Thus, $\widetilde{z}^{(i)}_{P,t}$
combines the displacement of the idiosyncratic component from its
long-run mean with the stock-specific drift.

In our application, we estimate a total of 15 market factors.\footnote{To handle missing price observations, we forward-fill return data, which is equivalent to the standard assumption that when prices are unobserved, they remain fixed to the last observation. In total, the factors we use account for more than 90\% of the variability of observed returns.} 
We build a total of six signals (stacked in vector $Z^{(i)}_t$) that differ in the level of discretization $P$. 
A core signal is computed following \eqref{eq:mod-z-score_signal} and setting  $P = 10$. We additionally built five alternative signals, also following \eqref{eq:mod-z-score_signal}, which differ from the core signal by setting $P \in \{20, 30, 60, 100, 150\}$.

\subsection{Return Prediction Problem}
\label{subsection:linear_model}

Once signals are constructed, we can set up the return prediction problem.
Under the simplifying assumption that the most recent signal observation contains sufficient information for forecasting, we have the general prediction equation
\begin{equation}\label{eq:general_return_pred}
    r^{(i)}_{t+h} = g_{t,h}\big( Z^{(i)}_t \big) + \epsilon^{(i)}_{t+h} ,
\end{equation}
where $g_{t,h} : \Real^D \to \Real$ is time and horizon-varying, possibly nonlinear function and $\epsilon^{(i)}_{t+h}$ is an error term. This setup is extremely broad, and estimation of $g_{t,h}$ is generally infeasible without additional restrictions on its functional form.

As a prediction baseline, we can consider a direct \textit{linear} forecasting model, where future returns are predicted as an affine function of signals,
\begin{equation}\label{eq:linear_return_pred}
    r^{(i)}_{t+h} = \mu_{t,h} + \beta_{t,h}' Z^{(i)}_t + \epsilon^{(i)}_{t+h} ,
\end{equation}
where $\mu_{t,h}$ is the market's average return, $\beta_{t,h} \in \Real^{D}$ is the vector of feature coefficients, which is, in general, horizon and time-dependent. We make the standard assumption that $\beta_{t,h}$ does not vary along the cross-section, that is, the signals we construct correctly embed any stock-specific information. 
Importantly, thanks to the linear structure of \eqref{eq:linear_return_pred}, $\mu_{t,h}$ and $\beta_{t,h}$ can be readily estimated by least-squares, yielding $\widehat{\mu}_{t,h}$ and $\widehat{\beta}_{t,h}$, respectively. The linear return prediction is thus $\widehat{r}^{(i)}_{t+h} = \widehat{\mu}_{t,h} + \widehat{\beta}_{t,h}' Z^{(i)}_t$. 

More sophisticated nonparametric and machine learning approaches may also be applied to estimate \eqref{eq:general_return_pred}. Among these, common choices in the time series econometrics literature are kernel and series-based estimators \citep{bosqNonparametricStatisticsStochastic1996,fanNonlinearTimeSeries2003, liNonparametricEconometricsTheory2007}, as well as neural networks, random forests, and boosting methods \citep{fransesNonlinearTimeSeries2008,masiniMachineLearningAdvances2023}.
To also provide a machine-learning-based benchmark, in our empirical exercise in Section~\ref{section:application}, we use random forests (RFs) to construct a nonlinear estimate of \eqref{eq:general_return_pred}.
Random forests were introduced by \cite{breimanRandomForests2001}, and have since become a popular machine learning approach; see \cite{biauRandomForestGuided2016} for a general introduction, 
\cite{scornetTheoryRandomForests2026} for theoretical properties, and \cite{fallahgoulHighDimensionalLearningFinance2025} for a recent discussion of applications in finance.

\section{Echo State Networks}\label{section:esn_models}

The idea that random parameters may work well in complex nonlinear models such as artificial neural networks, in fact, goes back a long way \citep{lowe1988multivariableFunctionalInterp,Schmidt1992feedforwardNeuralRandom}. 
Different research fields have developed this concept:
\cite{huang2006universalApproxRandom} proposed the ``extreme learning machine'' form for shallow NNs with randomly sampled internal weights;\footnote{By \textit{shallow} we intend here that the number of hidden layers of the network is small, usually only one. However, neural networks with even a single layer are effective functional approximators. Textbook discussions of such results can be found, e.g., in \cite{hastie2009theElements}, and also in \cite{anthony2009neuralNetwork} or \cite{zhang2023mathematical} for formal theory.}
\cite{rahimi2008UniformApproximationFunctions,rahimi2008randomKitchenSinks} developed the ``random kitchen sinks'' approach (i.e., collections of functions with random parameters) in the setting of reproducing kernel Hilbert spaces (RKHS). 
More recent work on pinpointing the reasons behind the success of deep neural networks has also looked into the ``lottery ticket'' hypothesis \citep{frankleLotteryTicketHypothesis2018,malachProvingLotteryTicket2020,maSanityChecksLottery2021}, which informally states that in large NNs there is a high chance that a small subnetwork performs almost as well as its much larger super-network---even at initialization, \textit{before} coefficients are trained \citep{sreenivasanRareGemsFinding2022}. Works such as  \cite{zhaoZerOInitializationInitializing2022} and \cite{bolagerSamplingWeightsDeep2023} go further and design special sampling schemes that try to optimize the initial weight sampling strategy, so that negligible-to-zero training is needed.
In tandem, \textit{reservoir computing} (RC) models have been developed as a (broad) family of methods which \textit{explictly} embrace that idea that most model parameters do not need to be trained, but can instead be randomly drawn from some distributions chosen by the researcher, independently of the data \citep{lukoseviciusPracticalGuideApplying2012}.\footnote{We remark that, while the coefficients are sampled at random, this does not mean that the models' \textit{architecture} is also necessarily random. Indeed, taking into account time dependencies is essential.} 
Reservoir models, more specifically, originate from the fields of dynamic system analysis and control theory.
An overarching subfamily of interest is thus that of Random Weight Neural Networks (RWNN), which encompasses all NN-like model designs where at least some of the weights are drawn randomly and not trained.

As mentioned in the introduction, an important consideration about RC and RWNN methods is their \textit{computational efficiency}. Artificial neural network training often demands significant, if not extreme, investments related to both physical resources (computer clusters, hardware, etc.) and energy; these costs have been steadily rising over the past years as models quickly become more complex \citep{Lohn2022aiComputeHowMuchLonger}. Since RC and RWNN models do not require full parameter training, the costs to optimize such models can be orders of magnitude lower. This also means that they can be effectively deployed in contexts where systematic re-training (e.g., online learning) is required.

\subsection{ESN Modeling Framework}

Echo State Networks are a relatively recent development in the field of nonlinear time series modeling. They belong to both reservoir computing and random weight network families. They are architecturally analogous to recurrent neural networks (RNN), in that they explicitly model the time-dependence of data using a state equation.

Formally, suppose that time series $Y_t$ and $Z_t$ are observed, and we seek to predict $Y_{t+1}$ using $Z_t$ and its past realizations (history). In our setting, $Z_t$ is a vector of stock-specific signals, while $Y_{t+1}$ is a stock return realized at a future horizon (e.g., 10-minute ahead).
Let $Z_t = (Z_{1,t}, \ldots, Z_{D,t} )' \in \Real^D$, where $D \geq 1$ is the number of stock signals, and $Y_t \in \Real$.
An ESN model maps features $Z_t$ to a \textit{recurrent state} $X_t$ defined by the equation
\begin{equation}\label{eq:esn_state}
    X_t := \alpha X_{t-1} + (1-\alpha) \varphi\big( A X_{t-1} + C Z_t + b \big) ,
\end{equation}
where $X_t \in \Real^K$, for $K \geq D$; $A \in \Real^{K \times K}$, $C \in \Real^{K \times D}$ and $b \in \Real^{K}$ are the state equation coefficients; $\varphi : \Real \mapsto \Real$ is a nonlinear, element-wise activation function (e.g. hyperbolic tangent or ReLU), and $\alpha \in [0,1]$ is the state leak hyperparameter. Note that in the special setting where $K = D$, $\varphi(x) = x$, $A = 0_D$ is the $D\times D$ zero matrix, $C = I_D$ is the identity matrix of size $D$, $b = 0$ is the zero vector, and $\alpha = 0$, the above state equation simplifies to $X_t \equiv Z_t$. Therefore, by introducing state $X_t$ we are considering a strictly more general class than that of linear models. 
In order to perform regression, the state must be mapped to an output value. The output of the ESN is determined by a linear layer,
\begin{equation}\label{eq:esn_output}
    X_t \:\mapsto\: \mu_t  +  \theta_t' X_t,
\end{equation}
where $\theta_t \in \Real^{K}$ are the output coefficients and $\mu_t \in \Real$ is the output intercept term. Notice here that, in our ESN construction, we explicitly allow for the output coefficients to depend on time index $t$, while state parameters ($A$, $B$, and $\alpha$) are constant.
Thanks to the linear specification of \eqref{eq:esn_output}, the model can be fitted by regularized least squares by solving
\begin{equation}\label{eq:esn_least_squares}
    \big\{ \widehat{\mu}_t, \widehat{\theta}_t \big\}
    \: := \: 
    \arg\min_{\mu, \theta} 
    \sum_{t=T_1}^{T_2} \left(
        Y_{t+1} - \mu - \theta' X_t 
    \right)^2 + \phi(\theta) ,
\end{equation}
where $\phi(\theta) > 0$ is the regularization term---e.g. $\phi(\theta) = \lambda \norm{\theta}_2^2$ for ridge regression with an isotropic penalty---and we have assumed that a data sample over periods $T_1 \leq t \leq T_2$, $T_1 < T_2$, is available.

\begin{remark}
    One can allow for more general state and output equations, and RC models with, for example, polynomial state equations and nonlinear output equations have been studied in the literature, c.f. \cite{gononApproximationBoundsRandom2023}. One issue with changing \eqref{eq:esn_output} to be nonlinear is that training can become much more complex. For example, letting the output layer be $X_t \mapsto g_{\theta}(X_t)$ where $g_{\theta} : \Real^K \to \Real$ is a neural network with parameters $\theta$ means that the ESN model will need to be trained by costly numerical optimization routines, e.g. stochastic gradient descent and its variants; hyperparameter tuning also becomes much more elaborate.
    Instead, we solve \eqref{eq:esn_least_squares} explicitly with cross-validated, ridge-regularized least squares (see Section~\ref{subsection:training} below).
\end{remark}

\begin{remark}
    In recent years, quantitative results on the approximation properties of ESNs and, more generally, reservoir computing models have been developed.
    \cite{grigoryeva2018EchoStateNetworksUniversal} and \cite{gonon2020ReservoirComputingUniversality} have shown that ESNs are universal approximators of time filters acting on, respectively, deterministic sequences and stochastic processes.
    In the stochastic setting more specifically, \cite{gononApproximationBoundsRandom2023} established that, if $Y_{t+1} = H^*(\bm{Z}_t)$, where $\bm{Z}_t := (Z_t, Z_{t-1}, \ldots)$ and with $H^* : (\Real^D)^{\Int_{-}} \to \Real$ satisfying appropriate regularity conditions, there exists an ESN such that the $L^2$ approximation error $\E[(Y_t -  H(\bm{Z}_t))^2]^{1/2}$ is of order $\mathcal{O}(1/\sqrt{K})$. This order of convergence is notably independent of the input dimension, $D$.
    Moreover, in \cite{gonon2020RiskBoundsReservoir} the same authors also developed generalization bounds for the out-of-sample error of ESN models.
\end{remark}

\begin{remark}
\label{remark:esn_vs_rnn}
    While ESNs can be readily related to classical RNNs, there is not yet a universal framework to formally compare the two approaches. This is in part due to the fact that {quantitative} approximation theory for RNN-type models is still developing; see \cite{wangInverseApproximationTheory2023a,jiaoApproximationBoundsRecurrent2025}. 
    From an applied perspective, ESNs and RNNs seem to have domain-dependent strengths and weaknesses, cf. \cite{vlachasBackpropagationAlgorithmsReservoir2020}. 
    It should also be noted that training of vanilla recurrent networks via gradient descent methods is known to be unstable, leading to the development of more robust (but not equivalent) alternatives, such as GRUs and LSTMs. Indeed, plain RNNs suffer greatly from exploding gradients during training \citep{hochreiterVanishingGradientProblem1998}.
    This latter concern in particular motivates our comparison with random forests rather than a fully-trainable recurrent network in Section~\ref{section:application}.
\end{remark}

We underline here that simply drawing $A$, $C$, and $b$ from some prescribed distribution will generally not yield an effective ESN model. This is because specific properties, such as the so-called \textit{echo state} and \textit{fading memory} properties \citep{grigoryeva2018EchoStateNetworksUniversal}, are essential in guaranteeing the well-posedness and effectiveness of reservoir models. 
Therefore, we introduce additional hyperparameters, following the discussion in e.g. \cite{Ballarin2024reservoirComputingMacro}.
One begins by drawing random matrices $A^*$, $C^*$, and $b^*$. These are then normalized according to a specific matrix (pseudo)norm, $\norm{\cdot}_A$, $\norm{\cdot}_C$ and $\norm{\cdot}_b$, respectively, yielding $\overline{A} := A^* / \norm{A^*}_A$, $\overline{C} := C^* / \norm{C^*}_C$ and $\overline{b} := b^* / \norm{b^*}_b$. The state equation then becomes
\begin{equation}\label{eq:esn_state_explicit}
    X_t := \alpha X_{t-1} + (1-\alpha) \varphi\big( \rho \overline{A}\, X_{t-1} + \gamma \overline{C}\, Z_t + \zeta \overline{b}\, \big) ,
\end{equation}
where $\rho \in [0,1]$, $\gamma > 0$, and $\zeta \geq 0$ are the spectral radius, input scaling, and bias scaling, respectively. The reason why $\rho$ is taken to be within the interval $[0,1]$ is that $\norm{\cdot}_A$ is most often chosen to either be the 2-norm or the largest absolute eigenvalue norm. We use the latter, so that the spectral radius of $\overline{A}$ is unity. 

\begin{remark}
    The way of sampling $A^*$, $C^*$, and $b^*$ is itself a question of high practical importance. Many works have addressed this topic from an applied or computational perspective; see, e.g., \cite{lukoseviciusPracticalGuideApplying2012} and references within. 
    From a theoretical perspective, \cite{gononApproximationBoundsRandom2023} derived explicit distributions for the state parameter matrices that depend on the structure of the approximation problem.
    In this work, we favor a practical perspective and apply some well-studied probability distributions. The entries of $A^*$ are drawn from a sparse Gaussian distribution, while the entries of $C^*$ are drawn from a sparse uniform distribution, since some empirical success with these has been observed in \cite{Ballarin2024reservoirComputingMacro}. Moreover, in all of our ESN models, $b$ is set to be the zero vector, without any random draw. Accordingly, we do not discuss the tuning of $\zeta$ in our application.
\end{remark}

\subsection{Return Prediction with ESN}
\label{subsection:training}

\begin{figure}[t]
    \centering
    \includegraphics[width=0.9\linewidth]{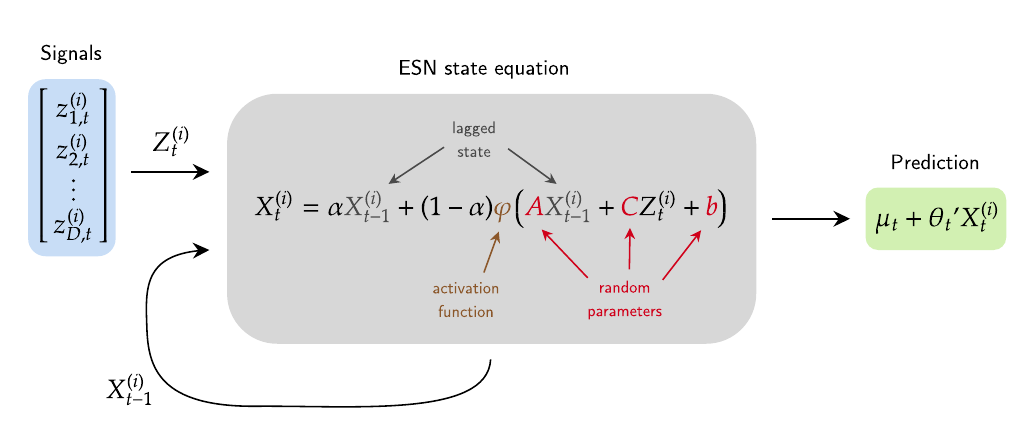}
    \caption{Overview of the proposed ESN model.}
    \label{fig:diagram_resn}
    \caption*{\small
    \textit{Notes:} The signal vector is loaded into the ESN state equation and combined with the previous state via a nonlinear mapping with randomly sampled parameters. The prediction equation is linear in the ESN states, allowing for (regularized) least squares estimation. The state parameters do not change when fitting the model, and only $\mu_t$, $\theta_t$ are estimated.}
\end{figure}

We now specialize the general ESN model from the previous section to the return prediction setting of interest. We provide a diagram of the flow of information in our ESN model in Figure~\ref{fig:diagram_resn}.
Following Section~\ref{subsection:linear_model}, we do not instantiate $N$ stock-specific ESN state equations, but rather draw a single set of state parameters that are then {shared} across stocks. 
Formally, we let the ESN state for the $i$th asset, $X^{(i)}_t$, be given by
\begin{equation*}
    X^{(i)}_t = \alpha X^{(i)}_{t-1} + (1-\alpha) \varphi\big( \rho \overline{A}\, X^{(i)}_{t-1} + \gamma \overline{C}\, Z^{(i)}_t \big) ,
\end{equation*}
where $\varphi$, $\alpha$, $\overline{A}$, $\overline{C}$, $\rho$ and $\gamma$ are fixed and identical across $i \in \{1, \ldots, N\}$. 
The stock-specific ESN regression equation thus becomes
\begin{equation*}
    r^{(i)}_{t+h} = \mu_{t,h} + \theta_{t,h}' X^{(i)}_t + \varepsilon^{(i)}_{t+h} .
\end{equation*}
Observe that $\mu_{t,h}$ and $\theta_{t,h}$ are also shared across stocks. As with the linear model, here too we assume that returns can be predicted from signals with regression coefficients that are common across the stock cross-section.
Since returns are, in general, overlapping and we employ a direct prediction strategy, the same state $X^{(i)}_t$ is used for all horizons.
To handle missing data, we introduce what we term \textit{state decay}, meaning that we replace missing signals with zeros, and continue iterating the state equation forward---although the associated state is not included in the estimated data; see Appendix~\ref{appendix:estimation_details} for details.

\begin{remark}
    Our choice to implement state decay instead of opting for data in- and back-filling operations, such as linear interpolation between missing values, is primarily due to two key concerns. 
    First, since we work with overlapping returns and online model estimation, we want to avoid information leakage from future observations to the past. 
    Secondly, the missingness patterns are very heterogeneous across individual stocks, sometimes with large gaps. 
    This makes it hard to construct valid and robust infill strategies \citep{bryzgalovaMissingFinancialData2025}, which may even have negative effects on performance \citep{chenMissingValuesHandling2024}.
\end{remark}

To train the ESN, our objective is to sequentially minimize the regularized least squares empirical risk,
\begin{equation}\label{eq:loss_estimation}
    \mathcal{R}_{t,h} (\mu, \theta)
    := 
    \frac{1}{M_{t,h}} \sum_{s=t-\tau_h-M_{t,h}}^{t-\tau_h-1} \frac{1}{N_s} \sum_{i \in \mathcal{I}_s} \left(r^{(i)}_{s+h} - \mu - \theta' X^{(i)}_s \right)^2 + \theta' \Lambda_{t,h} \theta ,
\end{equation}
where $1 \leq M_{t,h} \leq t-\tau_h$ is the \textit{training window size} for the current horizon, $0 \leq \tau_h \leq t-1$ is the \textit{training window buffer}, and $\Lambda_{t,h}$ is a positive semi-definite diagonal penalization matrix.%
Notice that $\mathcal{R}_{t,h} (\mu, \theta)$ is not written explicitly as a function of ESN state coefficients, as these are \textit{not} subject to training/fitting.
The ridge-type regularization penalty is encoded in the time-varying, horizon-specific matrix $\Lambda_{t,h}$. 

To operationalize \eqref{eq:loss_estimation}, $M_{t,h}$, $\tau_h$ and $\Lambda_{t,h}$ need to be set. 
The window size controls how far back the model can go to extract information from past realized signals. A small window means that a short panel of observations is utilized to optimize $\mathcal{R}_{t,h} (\mu, \theta)$, while a longer panel may include irrelevant information with respect to current market conditions. Therefore, we tune $M_{t,h}$ for each prediction horizon $h$ in our empirics. 
The window buffer $\tau_h$ is necessary because we want to ensure that in our training, we consider the time structure of multi-period returns. If $h = \textnormal{10 min}$, then at time $t$ we may fit the model using states starting from up to ten minutes ago; instead, if we consider $h = \textnormal{60 min}$, then clearly we can only construct training tuples $\{X^{(i)}_s, r^{(i)}_{s+h}\}$ where $s \leq t-h$, meaning the most recent states we can use are from time $t-\textnormal{2hr}$. If we want to avoid constructing an empirical estimation scenario where there is information leakage across time, imposing $\tau_h \geq h$ is necessary. Note that this implies that for horizon \textsc{eod} we are forced to use only observations up to the \textit{previous trading day}, which are valid for \textsc{eod} returns. These considerations need to be taken into account when considering the change in predictive performance of all methods, as presented in Section~\ref{section:application} below.
Lastly, the optimal ridge penalty could change substantially across both time and horizon. We use cross-validation to select $\Lambda_{t,h}$ in our empirical analysis, and we provide a detailed discussion of our implementation in Appendix~\ref{appendix:estimation_details}.

\section{Forecasting Multi-horizon Returns}\label{section:application}

\begin{table}[t]
	\centering
	\setlength\tabcolsep{5pt}
    \setlength\extrarowheight{1pt}
	\linespread{1}\selectfont\centering
    \caption{ESN Model Specifications}
	\begin{tabular*}{\textwidth}{@{\extracolsep{\fill}}*{7}{c}}
        & & \multicolumn{5}{c}{Horizon} \\
                \cline{3-7}
        \multicolumn{2}{c}{Parameter} & 10 min & 30 min & 60 min & 2 hours & \textsc{eod} \\
		\midrule
        \midrule
        $K$ & {\small State dim.} & 100 & 100 & 100 & 100 & 100 \\
        \midrule
        $\alpha$ & {\small Leak rate} & 0.9  & 0.2 & 0 & 0 & 0 \\
        \midrule
         & {\small $A$ sparsity} & 0.15 & 0.15 & 0.15 & 0.65 & 0.35 \\
        \midrule
        $\rho$ & {\small Spectral radius} & 0.4 & 0.6 & 0.6 & 0.6 & 0 \\
        \midrule
         & {\small $C$ sparsity} & 0.95 & 0.55 & 0.75 & 0.85 & 0.25 \\
        \midrule
        $\gamma$ & {\small Input scaling} & 0.005 & 0.005 & 0.005 & 0.005 & 0.015 \\
		\bottomrule
	\end{tabular*}
    \vspace{0.5em}
	\caption*{\small
    \textit{Note:} 
        All hyperparameters have been selected based on optimizing MSFE predictive performance on the Q4 2012 data sample.
    }
	\label{table:esn_specs}
\end{table}

We carry out the forecasting experiment using a period of time (i.e., 12 months of data) which, although short, is still performance-representative, especially because we are dealing with higher frequency data. Therefore, one calendar year still involves a large number of data points (16000) within the associated sample. The total number of stocks included in our dataset over this period is $N = 1290$. 
All forecasts are made in rolling-window fashion, so that at any given time index $t$, past observations are used to fit (and tune) the models, which then predict returns for all horizons $t+h$ with $h \in \{$10 min, 30 min, 60 min, 2 hours, \textsc{eod}$\}$.

\paragraph{ESN Models.}
To adapt our ESN model to each prediction horizon, our model specifications are determined as follows. 
The state dimension is fixed to $K=100$ for all horizons. We also set the random seed to a fixed value, so that we always draw the same $A^*$ and $C^*$, but vary hyperparameters $\rho$, $\gamma$, and $\alpha$, as well as the sparsity degree for the entries of both matrices. We collect our specifications in Table~\ref{table:esn_specs}. The final hyperparameters are selected by using a pre-sample from Q4 2012, over which we minimize the mean squared forecasting error at each specific return horizon.
In Appendix~\ref{appendix:estimation_details}, we provide more details on the numerical implementation of the hyperparameter optimization problem.

\paragraph{Linear Models.}

We employ the linear-in-signals and random forest models mentioned in Section~\ref{subsection:linear_model} as benchmarks for prediction.
In our experiments, we make a further modeling distinction between a ``simple'' linear model estimated via ordinary least squares and a \textit{ridge-regularized} linear model for predicting returns. 
Indeed, the linear specification is distinct from the particular estimation strategy, meaning that ridge regularization is a separate refinement. By also using a rolling batch, it may be possible to better capture temporary changes in the return and signal structure, include them in the penalization procedure, and thus improving prediction performance compared to a more na{\"i}ve least squares approach. 

We let the basic linear model be defined as in \eqref{eq:linear_return_pred}, and estimated its coefficients using OLS on the most recently observed signals (i.e., fixing the window size to $M_{t,h} = 1$). The risk thus simplifies to
\begin{equation*}
    \mathcal{R}_{t,h}^{\textnormal{linear}} (\mu, \beta)
    := 
    \frac{1}{N_t} \sum_{i \in \mathcal{I}_t} \left( r^{(i)}_{t-\tau_h-1+h} - \mu - \beta' Z^{(i)}_{t-\tau_h-1} \right)^2 .
\end{equation*}
The linear ridge regression model is estimated similarly to the ESN model by considering
\begin{equation*}
    \mathcal{R}_{t,h}^{\textnormal{ridge}} (\mu, \beta)
    :=
    \frac{1}{M_{t,h}} \sum_{s=t-\tau_h-M_{t,h}}^{t-\tau_h-1} \frac{1}{N_s} \sum_{i \in \mathcal{I}_s} \left(r^{(i)}_{s+h} - \mu - \beta' Z^{(i)}_{s} \right)^2 + \beta' \Lambda_{t,h} \beta , 
\end{equation*}
where the window size $M_{t,h}$ is set identically to the respective values for ESN fitting. In both cases, the window buffer $\tau_h$ is also chosen to be the same as in the ESN estimation. The penalty matrix $\Lambda_{t,h}$ is also selected using the same cross-validation scheme employed for the ESN model.
As discussed below, in our experiments, we find that baseline and benchmark linear models have very close performance. This suggests that we can primarily assign the improved performance to the nonlinear recurrent design of our ESN model, not the ridge regularization included in the estimation problem.

\paragraph{Random Forest.}
To further evaluate the impact of nonlinearities, as mentioned in Section~\ref{subsection:linear_model}, we also include a random forest (RF) regression model \citep{breimanRandomForests2001}. In our setting, RF regression aims to recover a nonlinear function $z \mapsto g_{t,h}(z)$ of the extracted signals defined as in equation \eqref{eq:general_return_pred}. 

Random forests are a stochastic ensemble of regression trees, which are conceptually similar to our chosen reservoir computing approach, since ESNs also exploit randomness. Formally, assuming $Z^{(i)}_{t} \in \mathcal{Z} \subseteq \Real^D$ for all $i$ and $t$, and letting $\{ B_j \}_{j=1}^J$ be a partition of $\mathcal{Z}$,\footnote{A collection of sets $\{ B_j \}_{j=1}^J$ is a partition of $\mathcal{Z}$ if $\cup_j B_j = \mathcal{Z}$ and $B_j \cap B_{j'} = \emptyset$ for any $j \not= j'$.} a regression tree is the local constant model
\begin{equation*}
    \psi\big( Z^{(i)}_{t} \big) 
    : =
    \sum_{j=1}^J b_j\ \1{ Z^{(i)}_{t} \in B_j } ,
\end{equation*}
where $\{b_j\}_{j=1}^J$ is the regression coefficient for partition element $B_j$.
A partition can be obtained by dividing the domain $\mathcal{Z}$ into cardinal-aligned rectangular regions by using a binary tree construction, and this is the approach taken by regression trees. 
RF regression is run by constructing $L > 1$ bootstrap resamplings of the original data, building a regression tree for each bootstrap sample, $\psi_\ell$ where $\ell = 1, \ldots, L$, and then averaging, yielding
\begin{equation*}
    \Psi\big( Z^{(i)}_{t} \big) 
    =
    \frac{1}{L} \sum_{\ell=1}^L \psi_\ell\big( Z^{(i)}_{t} \big)  .
\end{equation*}
The random forest prediction risk is thus given by
\begin{equation*}
    \mathcal{R}_{t,h}^{\textnormal{rf}} (\Psi)
    := 
    \frac{1}{N_t} \sum_{i \in \mathcal{I}_t} \left( r^{(i)}_{t-\tau_h-1+h} - \Psi\big( Z^{(i)}_{t-\tau_h-1} \big) \right)^2 ,
\end{equation*}
where once more we use the same window buffer $\tau_h$ as for ESN and linear models.
Random forests have been successfully applied to economic time series forecasting, although at present there is not yet a satisfactory theoretical analysis of their properties when applied to dependent data \citep{masiniMachineLearningAdvances2023}.
More broadly, tree-based and boosting methods have been shown to achieve state-of-the-art performance in with tabular (e.g., panel) data; see \cite{shwartz-zivTabularDataDeep2022,grinsztajnWhyTreebasedModels2022}.
We use the RF implementation of the \texttt{scikit-learn} Python package \citep{scikit-learn}; additional details on RF training are given in Appendix~\ref{appendix:estimation_details}.

\paragraph{Numerical Efficiency.}

One important concern when applying ML models in a high-frequency setting is how efficiently these can be estimated on broadly available hardware. 
We run all code on a system with a 64-core AMD EPYC 7763 CPU and 512GB of RAM; whenever possible, estimation routines are allowed to use all available cores for parallel computation.
In Table~\ref{table:runtime} we report the time necessary for each model to be fitted and yield forecasts over the entire sample.
It is easy to see that linear and ridge predictions are the most efficient models, followed by ESN, which is approximately $2.5$ times slower than the baseline linear regression.
The most expensive model by far is the random forest regression, which takes roughly $45$ times longer to run than the least squares forecasting scheme.
Frequent re-estimation of the random forest is significantly more expensive compared to the regularized estimation problem of the ESN.
The runtime comparison of Table~\ref{table:runtime} suggests that more complex ML models---such as fully-trainable feedforward/deep neural networks, or transformer-based time series foundation models \citep{jinLargeModelsTime2026}---would also be expensive to evaluate. For recurrent-type models, like RNNs, LSTMs, GRUs, and state-space models (SSMs), one faces the additional complexity due to training complexity (cf. Remark~\ref{remark:esn_vs_rnn}), missing observations, and possibly short time series for a large subset of stocks. 
In comparison, a core advantage of the proposed RC approach is that the ESN is both efficient in terms of computation and straightforward to apply, even with a few observations.

\begin{table}[t]
	\centering
	\setlength\tabcolsep{5pt}
    \setlength\extrarowheight{1pt}
	\linespread{1.1}\selectfont\centering
    \caption{Model Runtimes}
	\begin{tabular*}{\textwidth}{@{\extracolsep{\fill}}*{4}{c}}
		\toprule
        Linear & Ridge & RF & ESN \\
		\midrule
        \midrule
        56 seconds & 89 seconds & 2561 seconds & 139 seconds \\
        \small $1 \times$ & \small $1.58 \times$ & \small $45.73 \times$ & \small $2.48 \times$\\
		\bottomrule
	\end{tabular*}
    \vspace{0.5em}
	\caption*{\small
        \textit{Notes:} For ridge regression and ESN models, the reported time includes ridge penalty cross-validation. 
    }
	\label{table:runtime}
\end{table}

\subsection{Results}
\label{subsection:empirical_results}

First, we consider the mean squared forecasting error (MSFE). At prediction horizon $h$ and time $t$ at which we carry out the prediction, the \textit{instantaneous} MSFE across the $N$ stocks in the investable universe at time $t$ is given by
\begin{equation*}
    \textnormal{MSFE}_{t,h} 
    = 
        \frac{1}{N_t} \sum_{i = 1}^{N_t}(r^{(i)}_{t+h} - \widehat{r}^{(i)}_{t+h})^2 
    .
\end{equation*}
The lower the MSFE, the better the model's performance in a mean-squared error sense. Over an investment period, we also consider the cumulative MSFE, $\textnormal{cuMSFE}_{t,h} = \frac{1}{t} \sum_{s = 1}^{t} \textnormal{MSFE}_{s,h}$, as well as the total cumulated MSFE at terminal time $T$, that is $\textnormal{MSFE}_{h} \equiv \textnormal{cuMSFE}_{T,h}$.

\begin{figure}[p]
    \centering
    \caption{Relative Cumulative MSFE}
    \vspace{1em}
    \begin{subfigure}{0.49\textwidth}
        \includegraphics[width=\textwidth]{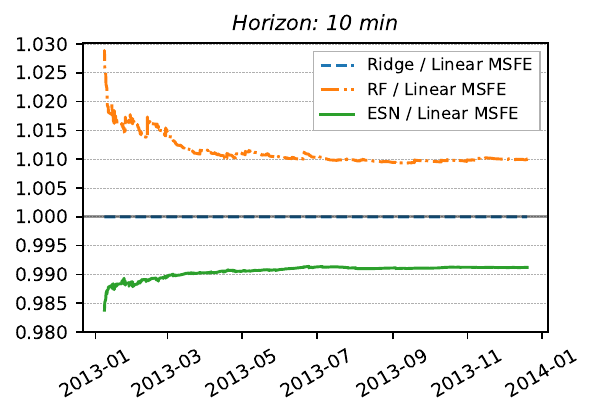}
        \label{fig:REL_CUM_MSFE_hr_10m}
    \end{subfigure}
    \hfill
    \begin{subfigure}{0.49\textwidth}
        \includegraphics[width=\textwidth]{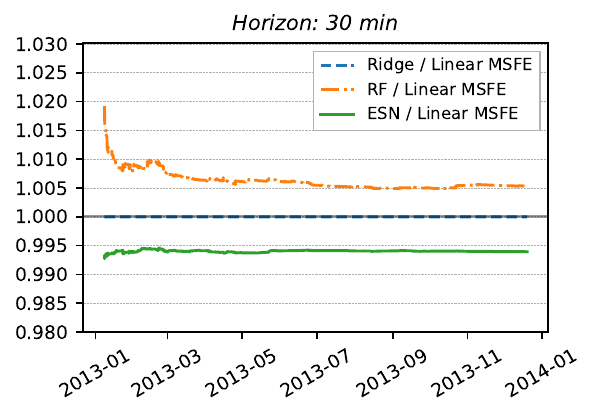}
        \label{fig:REL_CUM_MSFE_hr_30m}
    \end{subfigure} \\
    \vspace{1em}
    \begin{subfigure}{0.49\textwidth}
        \includegraphics[width=\textwidth]{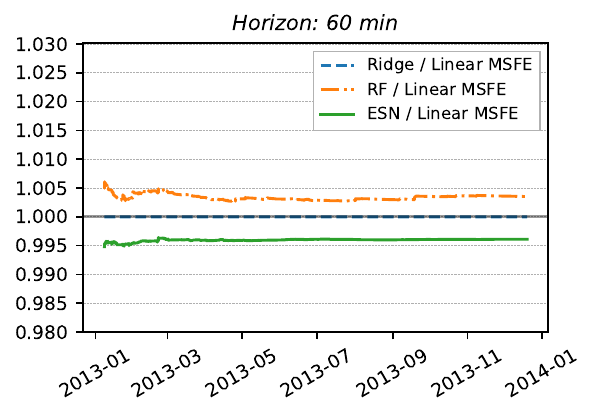}
        \label{fig:REL_CUM_MSFE_hr_60m}
    \end{subfigure}
    \hfill
    \begin{subfigure}{0.49\textwidth}
        \includegraphics[width=\textwidth]{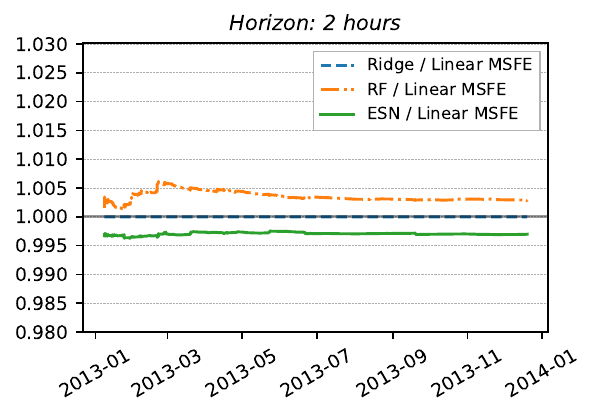}
        \label{fig:REL_CUM_MSFE_hr_2h}
    \end{subfigure} \\
    \vspace{1em}
    \hfill
    \begin{subfigure}{0.49\textwidth}
        \includegraphics[width=\textwidth]{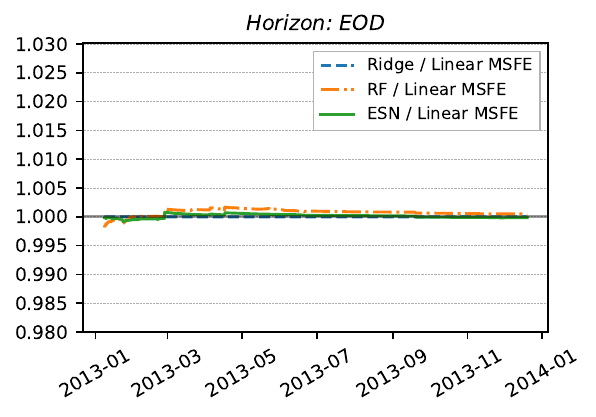}
        \label{fig:REL_CUM_MSFE_hr_eod}
    \end{subfigure}
    \hfill
    \vspace{1em}
	\caption*{\small
        \textit{Notes:} Relative cumulative mean squared forecasting error with respect to linear model baseline. The first week of January 2013 is omitted to avoid initial fluctuations.
    }
    \label{fig:REL_CUM_MSFE}
\end{figure}

Over the entire year of 2013, the cumulated MSFE of our ESN model is significantly lower than either the baseline linear or ridge regression models, as can be seen from Table~\ref{table:MSFE}. 
Interestingly, the random forest model is measurably worse than both linear benchmarks, by approximately the same margin as the ESN model in absolute terms.
In Figure~\ref{fig:REL_CUM_MSFE}, we also plot, for each of the five horizons we consider, the relative cumulative MSFE of all models---the baseline linear specification provides the reference point. 
In all cases, the difference between baseline and ridge is negligible: This is not exactly the case numerically, but MSFE deviations between the two models are insignificant for practical purposes. 
On the other hand, our ESN model is able to achieve a consistent reduction in MSFE of more than 0.87\% at the 10-minute horizon. 
However, this gain, while still robust over time, diminishes progressively as the horizon grows. In fact, the gain essentially vanishes when we consider \textsc{eod} returns, a setting where at times the cumulative error grows above that of the linear models. 
However, as evidenced in the last column of Table~\ref{table:MSFE}, even at \textsc{eod} returns, the ESN can extract a relative improvement that is two orders of magnitude larger than that of simply applying regularization to the linear baseline.

In Table~\ref{table:R2}, we also present the total $R^2$ of forecasts with respect to return realizations. It is easy to see that, while all values are negative, ESN predictions show improvement, especially at 10, 30, and 60-minute horizons.
We emphasize that, as \cite{kellyVirtueComplexityReturn2024} argue, negative $R^2$'s are not a sign of poor predictive or strategy efficacy. 
Accordingly, we find that, overall, our ESN framework attains better performance compared to our linear prediction benchmarks. 

\begin{table}[t]
	\centering
	\setlength\tabcolsep{5pt}
    \setlength\extrarowheight{1pt}
	\linespread{1}\selectfont\centering
    \caption{Total Cumulated MSFE}
	\begin{tabular*}{\textwidth}{@{\extracolsep{\fill}}*{6}{c}}
		\toprule
              & \multicolumn{5}{c}{Horizon} \\
                \cline{2-6}
        Model & 10 min & 30 min & 60 min & 2 hours & \textsc{eod} \\
		\midrule
        \midrule
        {Linear} & 0.0557 & 0.1402 & 0.2331 & 0.3704 & 0.7088 \\
        \midrule
        \multirow{2}{*}{Ridge} & 0.0557 & 0.1402 & 0.2331 & 0.3704 & 0.7088 \\
            & \small [-0.0010\%] & \small [-0.0007\%] & \small [-0.0004\%] & \small [-0.0003\%] & \small [-0.0001\%] \\
        \midrule
        \multirow{2}{*}{RF} & 0.0562 & 0.1409 & 0.2340 & 0.3714 & 0.7092 \\
            & \small [+1.0024\%] & \small [+0.5396\%] & \small [+0.3593\%] & \small [+0.2827\%] & \small [+0.0530\%] \\
        \midrule
        \multirow{2}{*}{ESN} & 0.0552 & 0.1393 & 0.2322 & 0.3693 & 0.7087 \\
            & \small [-0.8775\%] & \small [-0.6059\%] & \small [-0.3890\%] & \small [-0.3023\%] & \small [-0.0148\%] \\
		\bottomrule
	\end{tabular*}
    \vspace{0.5em}
	\caption*{\small
        \textit{Notes:} Percentage reduction in MSFE with respect to the linear model baseline is shown within squared brackets. Values are rounded to 4 decimal digits. 
    }
	\label{table:MSFE}
\end{table}

\begin{table}[t]
	\centering
	\setlength\tabcolsep{5pt}
    \setlength\extrarowheight{1pt}
	\linespread{1}\selectfont\centering
    \caption{Total Forecasting $R^2$s}
	\begin{tabular*}{\textwidth}{@{\extracolsep{\fill}}*{6}{c}}
		\toprule
              & \multicolumn{5}{c}{Horizon} \\
                \cline{2-6}
        Model & 10 min & 30 min & 60 min & 2 hours & \textsc{eod} \\
		\midrule
        \midrule
        {Linear} & -0.0766 & -0.1213 & -0.1128 & -0.1656 & -0.1760 \\
        \midrule
        {Ridge} & -0.0766 & -0.1213 & -0.1127 & -0.1656 & -0.1760 \\
        \midrule
        {RF} & -0.0870 & -0.1227 & -0.1167 & -0.1689 & -0.1766 \\
        \midrule
        {ESN} & -0.0675 & -0.1146 & -0.1084 & -0.1621 & -0.1758 \\
		\bottomrule
	\end{tabular*}
    \vspace{0.5em}
	\caption*{\small
    \textit{Note:} Values are rounded to 4 decimal digits.}
	\label{table:R2}
\end{table}

\paragraph{Performance Testing.}

Next, we argue that the gains enjoyed by our echo state network model are also significant from a statistical point of view.
To test for the statistical significance of these improvements, we perform both pairwise Diebold-Mariano (DM) tests \citep{diebold1995comparing,harvey1997testing} and Model Confidence Set (MCS) tests at all horizons \citep{hansenModelConfidenceSet2011}.

\begin{figure}[p]
    \centering
    \caption{Diebold-Mariano Tests}
    \vspace{1em}
    \begin{subfigure}{0.49\textwidth}
        \includegraphics[width=\textwidth]{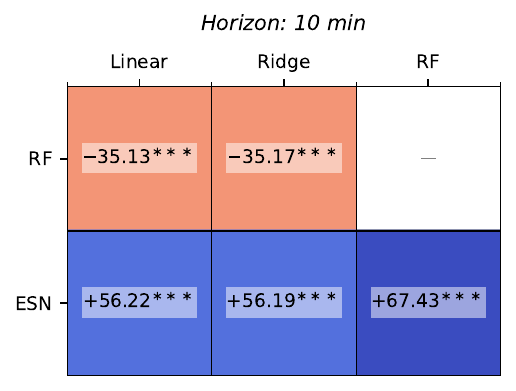}
        \label{fig:DM_test_hr_10m}
    \end{subfigure}
    \hfill
    \begin{subfigure}{0.49\textwidth}
        \includegraphics[width=\textwidth]{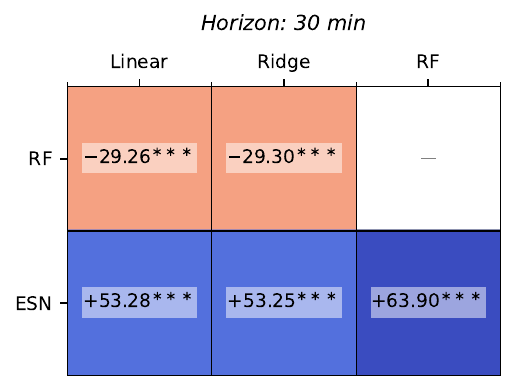}
        \label{fig:DM_test_hr_30m}
    \end{subfigure} \\
    \vspace{1em}
    \begin{subfigure}{0.49\textwidth}
        \includegraphics[width=\textwidth]{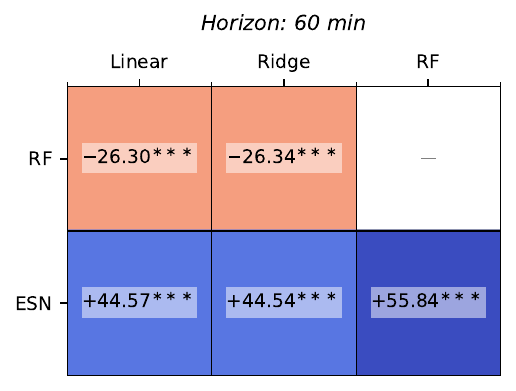}
        \label{fig:DM_test_hr_60m}
    \end{subfigure}
    \hfill
    \begin{subfigure}{0.49\textwidth}
        \includegraphics[width=\textwidth]{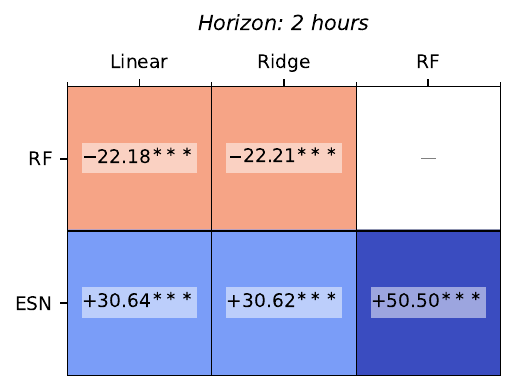}
        \label{fig:DM_test_hr_2h}
    \end{subfigure} \\
    \vspace{1em}
    \hfill
    \begin{subfigure}{0.49\textwidth}
        \includegraphics[width=\textwidth]{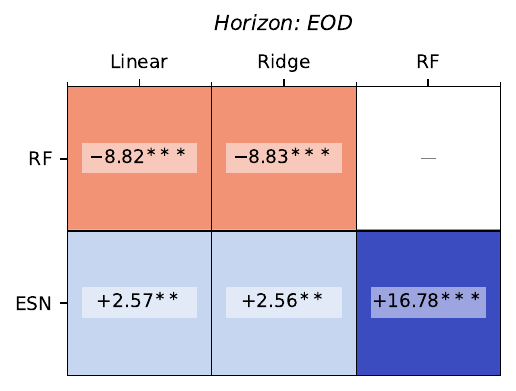}
        \label{fig:DM_test_hr_eod}
    \end{subfigure}
    \hfill
    \vspace{1em}
	\caption*{\small
        \textit{Notes:} Color matrix plots for one-sided Diebold-Mariano test statistics, column model versus row model. Statistical significance at 1\% ($^{***}$), 5\% ($^{**}$) and 10\% ($^{*}$) is shown next to statistic value.
    }
    \label{fig:dm_results}
\end{figure}

\begin{table}[t]
	\centering
	\setlength\tabcolsep{5pt}
    \setlength\extrarowheight{1pt}
	\linespread{1}\selectfont\centering
    \caption{Model Confidence Set Tests}
	\begin{tabular*}{\textwidth}{@{\extracolsep{\fill}}*{6}{c}}
		\toprule
              & \multicolumn{5}{c}{Horizon} \\
                \cline{2-6}
        Model & 10 min & 30 min & 60 min & 2 hours & \textsc{eod} \\
		\midrule
        \midrule
        Linear  & 0 & 0 & 0 & 0 & 0.0011 \\
        \midrule
        Ridge   & 0 & 0 & 0 & 0 & 0.0002 \\
        \midrule
        RF      & 0 & 0 & 0 & 0 & 0.5766$^{\checkmark}$ \\
        \midrule
        ESN     & 1$^{\checkmark}$ & 1$^{\checkmark}$ & 1$^{\checkmark}$ & 1$^{\checkmark}$ & 1$^{\checkmark}$ \\
		\bottomrule
	\end{tabular*}
    \vspace{0.5em}
	\caption*{\small
    \textit{Notes:} 
        Tests are run with the cross-sectional mean squared forecasting error using $10^4$ bootstrap draws. The checkmark superscript ($^{\checkmark}$) is used to indicate inclusion of the method into the set of best prediction models at the 5\% level.
    }
	\label{table:mcs_results}
\end{table}

For the Diebold-Mariano tests, we focus only on comparing our ESN model with the benchmark and baseline linear predictions. Following the test results summarized in Figure~\ref{fig:dm_results}, one can see that the null hypothesis of equal predictive ability is rejected extremely strongly---in particular, only at the end-of-day horizon, the $p$-values are above the 1\% level.
The MCS tests, on the other hand, enable comparison between all models simultaneously. As Table~\ref{table:mcs_results} shows, in all settings apart from \textsc{eod} returns, our ESN models robustly dominate; with end-of-day data, however, both the ESN and the RF model are included in the best models at a 5\% level. 
This outcome is not surprising, given the relative lack of improvement in cumulative MSFE shown in Figure~\ref{fig:REL_CUM_MSFE} for the \textsc{eod} horizon.

\subsection{Robustness to Parameter Sampling}
\label{section:robust_rand}

Since our Echo State Network prediction approach is inherently dependent on random draws of the state equation's matrices, we check whether different draws of our model yield substantially different forecasting performance. 
To achieve this, we focus on 10-minute and 60-minute horizons and sample 100 distinct ESN models---meaning we redraw $A^*$ and $C^*$ with different random seeds, while keeping unchanged the hyperparameters of Table~\ref{table:esn_specs}---and re-run our empirical exercise. We provide relative cumulative MSFE plots showing 90\% and 50\% frequency bands, as well as median performance.

\begin{figure}
    \centering
    \caption{ESN Model Robustness}
    \begin{subfigure}{0.49\textwidth}
        \includegraphics[width=\textwidth]{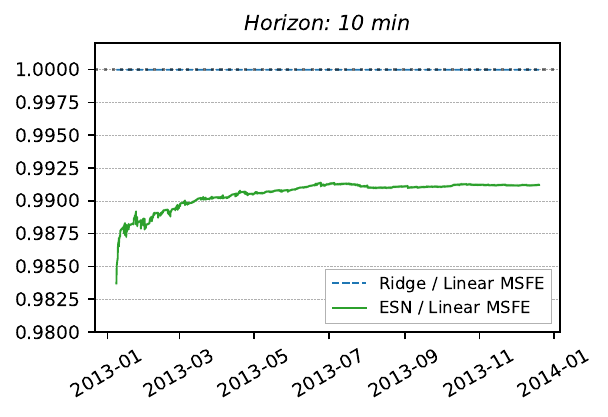}
    \end{subfigure}
    \hfill
    \begin{subfigure}{0.49\textwidth}
        \includegraphics[width=\textwidth]{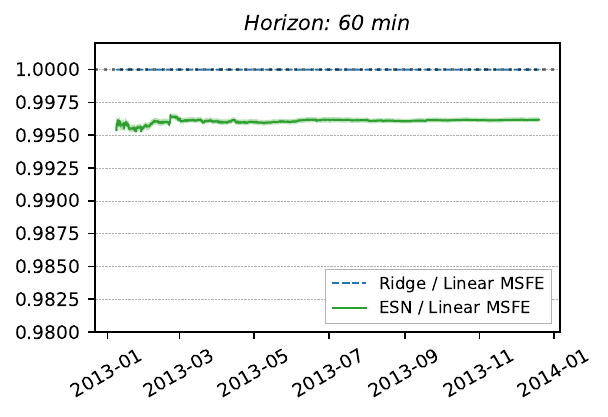}
    \end{subfigure}
    \vspace{5pt}
    \caption*{\small
    \textit{Notes:} Light and dark shaded bands show 90\% and 50\% frequency regions, respectively, for ESN models. Median ESN performance is shown as a solid line. Note that for both horizons, the bands are thin and MSFE is concentrated around the median. The first week of January 2013 is omitted to avoid initial fluctuations.}
    \label{fig:ROBUST_MSFE}
\end{figure}

From Figure~\ref{fig:ROBUST_MSFE}, we can see that the forecasting gains enjoyed by ESN models are remarkably robust to resampling state parameters.  In fact, while we do draw shaded regions for relative cumulative MSFE interquantile ranges, these are very tight and thus hard to see, indicating that predictive performance is consistent. In both horizons, the MSFE metric is highly concentrated and robust to ESN model sampling randomness. 
Thus, we confirm that our approach yields robust gains over benchmark prediction models.

\subsection{Portfolio Evaluation}

To evaluate the economic implications of our forecasts, we use them in the construction of a classical mean-variance portfolio over the same testing period \citep{roncalliIntroductionRiskParity2016}. We compare the ESN-based portfolio with benchmarks constructed using the linear, ridge, and random forests predictions, as well as a reference equal-weighted (EW) portfolio.%
For the sake of comparability, we subset the large panel of assets available to avoid including stocks that are thinly traded and, thus, may induce strictly mechanical changes in positions. This is particularly important when rebalancing at high frequencies. The final panel is has uniform cross section of size $\overline{N} = 388$. 
We focus on the 10-minute return horizon, as this setting does not require one to handle the time overlap that occurs at coarser horizons,\footnote{Intra-day overlapping returns at 10-minute frequency and hourly horizon, for example, require one to choose among 5 arbitrary starting times at the beginning of each day. Moreover, end-of-day returns do not allow for intra-day rebalancing, as \textsc{eod} returns are only observed at market close.} and shows the best improvement in MSFE for our ESN model; see Section~\ref{subsection:empirical_results}.
We provide a detailed discussion of asset selection and portfolio weight estimation in Appendix~\ref{appendix:portfolio}.

To determine the weights of portfolios based on return forecasts, we first require an estimate of the asset covariance matrix. As we only have 39 intraday realizations of 10-minute returns, the estimation problem is high-dimensional; moreover, since the data is high-frequency, we need to account for changes in correlation across days and weeks. Hence, we estimate the covariance matrix at a daily frequency using a nonlinear Ledoit-Wolf shrinkage estimator \citep{ledoitWellconditionedEstimatorLargedimensional2004}. The previous day's covariance estimate, $\widehat{\Sigma}_{\lfloor t \rfloor}$, is shared in the computation of the portfolio weights for all forecast methods, given by
$
    \widehat{\omega}_{t}^{m} = \widehat{\Sigma}_{\lfloor t \rfloor}^{-1} \, \widehat{r}_{t+\textnormal{10 min}}^{\,m}
$,
with $m \in \{ \textnormal{Linear, Ridge, RF, ESN} \}$. We normalize weights so that $\sum_{i=1}^{\overline{N}} = 1$, and we initialize all portfolios with equal weights.

\begin{figure}
    \centering
    \caption{Cumulative Portfolio Returns and Turnovers}
    \begin{subfigure}{0.49\textwidth}
        \includegraphics[width=\textwidth]{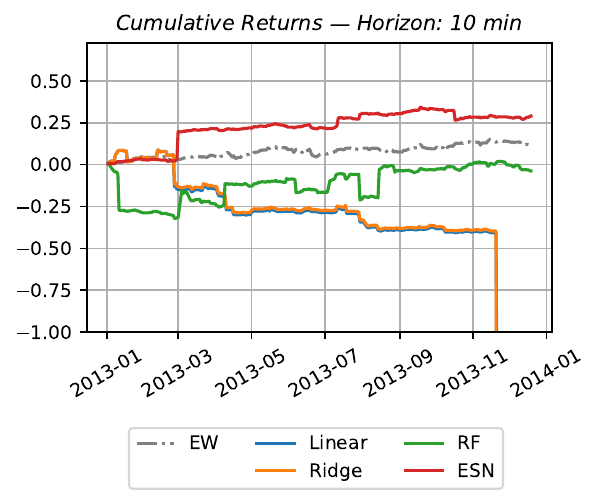}
    \end{subfigure}
    \hfill
    \begin{subfigure}{0.47\textwidth}
        \includegraphics[width=\textwidth]{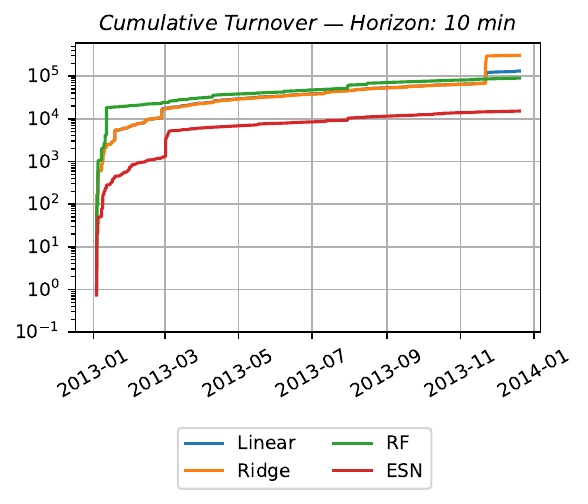}
    \end{subfigure}
    \vspace{5pt}
    \caption*{\small
    \textit{Notes:} Portfolio weights derived with linear, ridge, random forest, and ESN forecasts are computed on identical daily covariance matrix estimates with Ledoit-Wolf nonlinear shrinkage. Cumulative returns are net of transaction costs. Cumulative turnover is shown on a logarithmic axis. }
    \label{fig:portfolio_cumret_turnover}
\end{figure}

In Figure~\ref{fig:portfolio_cumret_turnover}, we show cumulative returns for all portfolios, and the cumulated turnover for forecast-based strategies.\footnote{To plot cumulative returns, we use a reduced range for the $y$-axis due to instabilities in the linear and ridge portfolios. Figure~\ref{fig:portfolio_cumret_full} in the appendix shows cumulative returns on an unrestricted axis.} 
We can see that the ESN-based portfolio performs best, achieving raw returns higher than the benchmark EW strategy. Linear, ridge, and RF forecasts, on the other hand, produce volatile positions due to constant and substantial weight changes (see Figure~\ref{fig:daily_average_weights_hr_10m} in the Appendix). This is evidenced by much higher cumulative turnovers induced by these methods compared to the one achieved by the ESN portfolio weights. 
We also report Sharpe ratios and maximum drawdowns in Table~\ref{table:sharpe_drawdown}. Among the forecast-based strategies, the ESN portfolio achieves the highest Sharpe ratio and the lowest maximum drawdown. Moreover, as Figure~\ref{fig:portfolio_cumret_turnover} shows, the ESN forecasts induce substantially lower portfolio turnover than the linear, ridge, and RF forecasts. An important distinction arises once transaction costs are taken into account, as the passive EW portfolio can outperform the actively rebalanced forecast-based strategies net of transaction costs (as shown in Figure~\ref{fig:portfolio_net_cumreturns} in the Appendix). 
We do not view this result as contradicting the forecasting gains documented above. Rather, it illustrates the important distinction between statistical predictability and the economic implementability of a forecasting signal. At the 10-minute rebalancing frequency considered here, even relatively small changes in predicted returns can generate substantial changes in the weights of an unconstrained mean-variance portfolio, and consequently large transaction costs. In this respect, the substantially lower turnover generated by the ESN portfolio is itself noteworthy. Relative to the other forecast-based strategies, the ESN moves considerably towards converting improved predictive accuracy into a more stable portfolio allocation. Whether these forecasting gains can be fully translated into superior net economic performance, however, depends also on the portfolio construction mechanism. A realistic implementation would typically incorporate transaction costs directly into the optimization problem or passive orders for some or all the assets \citep{ledoitMarkowitzPortfoliosTransaction2025}, and impose turnover, position, leverage, and other trading constraints. We deliberately do not introduce such constraints here, since our objective is to compare the economic implications of the different forecasting methods under the same simple portfolio mapping. Investigating the interaction between ESN forecasts and a fully constrained portfolio optimization framework is an important direction for future work.

\begin{table}[t]
	\centering
	\setlength\tabcolsep{5pt}
    \setlength\extrarowheight{1pt}
	\linespread{1.1}\selectfont\centering
    \caption{Portfolio Sharpe Ratios and Maximum Drawdowns}
	\begin{tabular*}{\textwidth}{@{\extracolsep{\fill}}*{6}{c}}
		\toprule
        & EW & Linear & Ridge & RF & ESN \\
		\midrule
        \midrule
        Sharpe ratio & 1.422 & $-1.181$ & $-1.044$ & 0.166 & 1.529 \\[4pt]
        Max. drawdown & $-0.743$ & $-22.882$ & $-85.350$ & $-33.057$ & $-0.465$ \\
		\bottomrule
	\end{tabular*}
    \vspace{0.5em}
	\caption*{\small
        \textit{Notes:} Sharpe ratios are computed using total daily returns and annualized. 
    }
	\label{table:sharpe_drawdown}
\end{table}

\section{Conclusion}\label{section:conclusion}

In this work, we have presented a novel approach to forecast intraday stock returns at different horizons by leveraging ESN models. Our proposed method is nonparametric in nature and capable of significantly improving prediction performance across different horizons by a significant margin, with the only exception being end-of-day returns.
Our findings suggest that there is potentially much to be gained by considering efficient, nonlinear forecasting models compared to regularized linear methods.
Even though our ESN models inherently rely on randomly sampled parameters, we have shown that the prediction gains are robust to randomness at both short and medium intraday horizons. 
This robustness extends to comparison with alternative machine learning models, such as the widely-used random forest regressor.
Additionally, we have established that ESN forecasting gains can translate to meaningful improvements in performance (compared, e.g., to equal-weighting schemes) when used to construct mean-variance portfolio strategies at a high-frequency, 10-minute horizon.

While it is beyond the scope of this work, one important thing we do not explore is whether our prediction method can also yield better performance---in terms of Sharpe ratio, for example---once applied specifically to \textit{multi-horizon} MV portfolio construction. We consider this an important direction along which to extend our work, which nonetheless also requires the derivation of a realistic framework to jointly model overlapping returns and their forecasts.
Market structure may also change frequently and suddenly, with potentially drastic changes in the volatility of and correlation between asset returns \citep{gottschalkBlackWednesdayBrexit2023}. This is something that is inderictly observed in our portfolio analysis, and we expect that taking into account this factor may help reduce portfolio turnover.
Moreover, as shown in our robustness analysis, it is straightforward to instantiate a large collection of diverse ESN models. Therefore, there is the possibility to combine our prediction framework with (online) model selection schemes, which can dynamically weight multiple ESNs to improve forecasting performance. 
For ESN models specifically, the ensemble approach was recently explored in \cite{ballarin2025fromManyModels}.  
Lastly, for the sake of simplicity and computational efficiency, we have constrained our ESN output layer to be linear. This is not a necessity, and it would be interesting to evaluate whether including more flexible output layers (e.g., a fully trainable shallow network) could further improve performance, and whether any gains obtained would be worthwhile given the increase in training complexity.

\newpage

\bibliographystyle{apalike}
\bibliography{rc_stock_returns.bib}

\newpage
\appendix

\counterwithin*{table}{section}
\renewcommand{\thetable}{\Alph{section}.\arabic{table}}
\counterwithin*{figure}{section}
\renewcommand{\thefigure}{\Alph{section}.\arabic{figure}}

\begin{center}\Large
    APPENDIX
\end{center}

\section{AlgoSeek Intraday Stock Data}\label{appendix:data}

We provide a summary snapshot of the data we have access to for each minute and each traded contract in Table~\ref{table:algoseek} below:

\bigskip

\begin{table}[h]
    \centering
    \renewcommand{\arraystretch}{1.3}
    \caption{AlgoSeek OHLC bar table example}
    \begin{tabularx}{\textwidth}{clp{3cm}X}
        \#  & \textbf{Field}            & \textbf{Type}        & \textbf{Description} \\ 
        \hline
        \hline
        1   & Date                      & YYYYMMDD             & Trade Date. \\ \hline
        2   & Ticker                    & String               & Ticker Symbol. \\ \hline
        3   & TimeBarStart               & HHMM,\newline HHMMSS,\newline HHMMSSMMM & 
        Time stamps are EST. For minute bars, the format is HHMM. For second bars, the format is HHMMSS. \newline
        Examples: \newline
        One second bar 130302 is from time greater than 130302.000 to 130302.999. \newline
        One minute bar 1104 is from time greater than 110400.000 to 110459.999. \\ \hline
        4   & FirstTradePrice            & Number               & Price of first trade. \\ \hline
        5   & HighTradePrice             & Number               & Price of highest trade. \\ \hline
        6   & LowTradePrice              & Number               & Price of lowest trade. \\ \hline
        7   & LastTradePrice             & Number               & Price of last trade. \\ \hline
        8   & VolumeWeightPrice          & Number               & 
        Trade Volume weighted average price: \newline
        \textit{Sum( (Trade1Shares$\times$Price)+(Trade2Shares$\times$ Price)} \textit{+ ... ) / TotalShares} \\ \hline
        9   & Volume                     & Number               & Total number of shares traded. \\ \hline
        10  & TotalTrades                & Number               & Total number of trades. \\ \hline
    \end{tabularx}
    \label{table:algoseek}
\end{table}

\section{Signal Construction}\label{section:signal_construction}

The signal construction workflow requires the estimation of market factors to obtain the residual idiosyncratic term $U^{(i)}_t$. 
We use PCA on the covariance matrix of returns to consistently estimate the factors:
\begin{itemize}
    \item We transform each stock return time series into a $z$-score, and subsequently calculate the associated pairwise covariance matrix as 
    \begin{equation*}
        \rho_{ij} = \frac{1}{T-1}\sum_{t=1}^{T} L^{(i)}_{t} L^{(j)}_{t}
        \quad\textnormal{and}\quad
        L^{(i)}_{t} = \frac{r^{(i)}_{t} - \overline{r}^{(i)}}{\overline{\sigma}^{(i)}} .
    \end{equation*}
    \item From the correlation matrix, PCA yields a vector of eigenvalues and corresponding eigenvectors $v_{i}^j$. For each index $j$, the eigenportfolio and the eigenportfolio returns are thus
    \begin{equation} %
        Q_{i}^j = \frac {v_i^j}{\sigma^{(i)}}
        \quad\textnormal{and}\quad
        F_{j,t} = \sum_{i=1}^{N}\frac {v_{i}^j}{\sigma^{(i)}}\, r^{(i)}_{t}, \quad j = 1,2,\ldots,J.
    \end{equation}
\end{itemize}

We can then proceed as discussed in Section~\ref{subsection:signals} to estimate the discrete-time analog of the OU process for $U^{(i)}_t$, which then yields, combining models at different levels of discretization, our vector of stock-specific signals.

\section{Model and Estimation Details}
\label{appendix:estimation_details}

\begin{figure}
    \centering
    \caption{Signal Availability over the Forecasting Period}
    \includegraphics[width=0.9\linewidth]{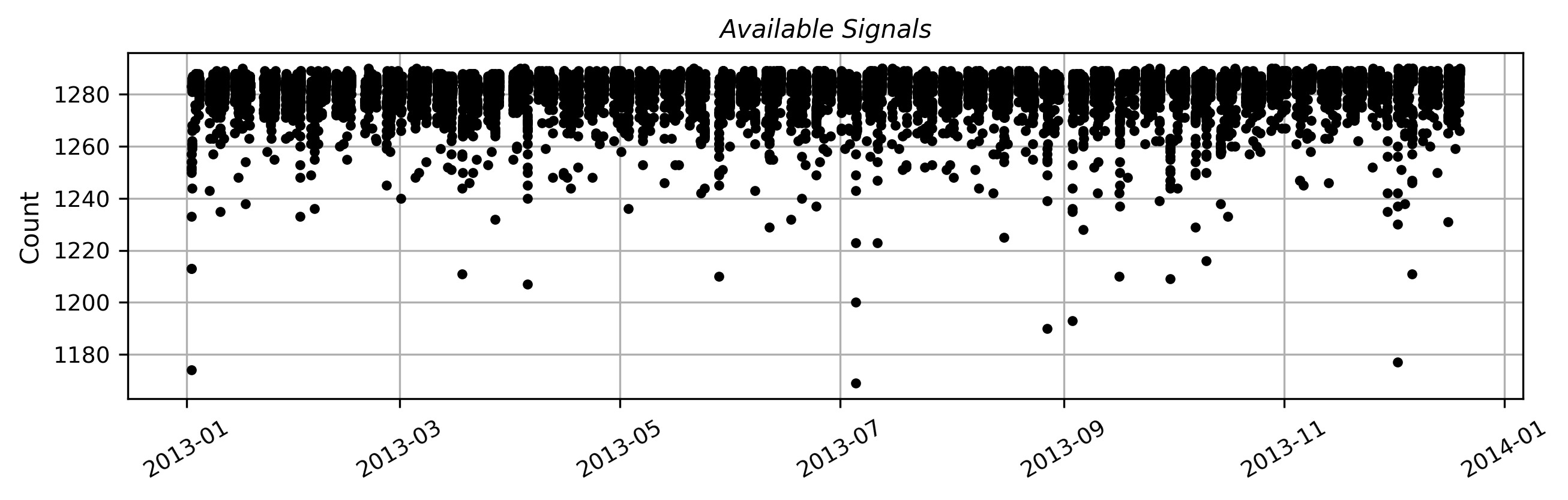}
    \label{fig:signal_avail}
\end{figure}

\paragraph{Missing Data and State Decay.}
As discussed in Section~\ref{subsection:data_setup}, the entire cross-section of returns and signals might not be fully available at all time indices. 
Indeed, as shown graphically in Figure~\ref{fig:signal_avail}, although consistently high, the count of observable asset signals changes often over the course of our panel.
As such, we design our ESN models in such a way as to be robust to missing data. Our strategy is to introduce \textit{reservoir decay} whenever input signals are missing.
Formally, recall that the $i$th stock state equation can be written as
\begin{equation*}
    X^{(i)}_t := \alpha X^{(i)}_{t-1} + (1-\alpha) \varphi\big( A X^{(i)}_{t-1} + C Z^{(i)}_t \big) .
\end{equation*}
Suppose now that the ESN must ingest a signal sequence that includes a \texttt{NaN} value,
\begin{equation*}
    \ldots,\: Z^{(i)}_{t-1},\: \texttt{NaN},\: Z^{(i)}_{t+1},\: \ldots
\end{equation*}
where we use \texttt{NaN} to signify missing data, as in the numerical setting. Reservoir decay functionally means that any \texttt{NaN} entry in the sequence of input signals is replaced with a zero vector. Therefore, we would obtain states
\begin{align*}
    X^{(i)}_{t-1} 
        & := \alpha X^{(i)}_{t-2} + (1-\alpha) \varphi\big( A X^{(i)}_{t-2} + C Z^{(i)}_{t-1} \big) , \\
    X^{(i)}_{t} 
        & := \alpha X^{(i)}_{t-1} + (1-\alpha) \varphi\big( A X^{(i)}_{t-1} \big) , \\
    X^{(i)}_{t+1}  
        & := \alpha X^{(i)}_{t} + (1-\alpha) \varphi\big( A X^{(i)}_{t} + C Z^{(i)}_{t+1} \big) .
\end{align*}
So long as $\alpha \in [0,1)$ and the spectral radius of $A$ is below unity, this approach allows treating sequences with arbitrary sub-sequences of missing data: The state equation is validly iterated forward, and the states decay (contract) towards the zero vector. This enables us not to ``throw away'' our previously collected state sequence.
Nonetheless, to conform with the setup of Section~\ref{subsection:data_setup}, any state that is associated with state decay induced by missing signal observation (or missing returns) is not included at the time of estimation for the output coefficients. 

\paragraph{Training Windows and Buffers.}
The training procedure outlined in Section~\ref{subsection:training} requires the choice of some parameters, such as training window size and buffer, as well as cross-validation frequency, window size, and ratio. We collect our choices for these quantities in Table~\ref{table:training_pars}.

\begin{table}[t]
	\centering
	\setlength\tabcolsep{5pt}
    \setlength\extrarowheight{1pt}
	\linespread{1}\selectfont\centering
    \caption{ESN Training and Cross-validation Parameters}
	\begin{tabular*}{\textwidth}{@{\extracolsep{\fill}}*{7}{c}}
        & & \multicolumn{5}{c}{Horizon} \\
                \cline{3-7}
        \multicolumn{2}{c}{Training Parameter} & 10 min & 30 min & 60 min & 2 hours & \textsc{eod} \\
		\midrule
        \midrule
        $M_{t,h}$ & {\small Window size} & 30 min & 30 min & 60 min & 60 min & 1 day \\
        \midrule
        $\tau_h$ & {\small Window buffer} & 10 min & 30 min & 60 min & 2 hours & 1 day \\
        \midrule
         & {\small CV frequency} & 1 day  & 1 day & 1 day & 1 day & 1 day \\
        \midrule
         & {\small CV window size} & 1 week & 1 week & 1 week & 1 week & 1 week \\
        \midrule
         & {\small CV split ratio} & 0.7 & 0.7 & 0.7 & 0.7 & 0.7 \\
		\bottomrule
	\end{tabular*}
	\label{table:training_pars}
\end{table}

For window size, we compromise between the length of the window and the negative effect that including too much past data can have on predictive power. For the window buffer, we always choose the minimal length possible given the prediction horizon.
Cross-validation to select the diagonal, anisotropic ridge penalty matrix $\Lambda_{t,h}$, is run every trading day with one week of past data. We keep a constant split ratio of $70\%$ training and $30\%$ validation across all horizons.

\paragraph{ESN Hyperparameter Optimization.}
To search for the horizon-specific ESN hyperparameters detailed in Table~\ref{table:esn_specs}, we rely on the Python model optimization library Optuna \citep{optuna2019}.
Our tuning sample includes only data from September to December 2012. We keep the same training and cross-validation parameters outlined in Table~\ref{table:training_pars}, and set as the objective to minimize the cumulated MSFE.

\paragraph{Random Forest Hyperparameters.}
We use \texttt{RandomForestRegressor} from \texttt{scikit-learn}. 
The following hyperparameters are set for all horizons: \texttt{min\_samples\_split=10}, \texttt{max\_depth=10}, and \texttt{n\_estimators=100}. 
To ensure that the RF is grown by bootstrapping trees also across different subsets of signals in $Z_t^{(i)}$, we set \texttt{max\_features='sqrt'}.
Finally, the option \texttt{n\_jobs=-1} is set to allow for parallelization across all available CPU cores. 

\section{Details on Portfolio Construction}
\label{appendix:portfolio}

\paragraph{Asset Selection.}
We construct a dense subset of assets based on volume at the 10-minute interval and 10-minute horizon, which in our setting is the only combination that ensures non-overlapping returns. We enforce that all assets included in the final panel used for portfolio construction and evaluation are such that both signals and returns are available at all 10-minute observation timestamps. This selection thus removes thinly traded or illiquid assets that may not be traded at high frequency. The final fixed cross-section size is $\overline{N} = 388$, which is still an order of magnitude larger than the number of daily observations.

\paragraph{Covariance estimation.}
To estimate the previous day's covariance estimate, $\widehat{\Sigma}_{\lfloor t \rfloor}$, we use the nonlinear shrinkage estimator of \cite{ledoitWellconditionedEstimatorLargedimensional2004} as implemented by the \texttt{scikit-learn} routine \texttt{LedoitWolf} with default block size.
This estimate is then used in the construction of all forecast-based MV portfolio weights to ensure comparability across methods.

\begin{figure}[p]
    \centering
    \caption{Portfolio Weights}
    \includegraphics[width=0.9\linewidth]{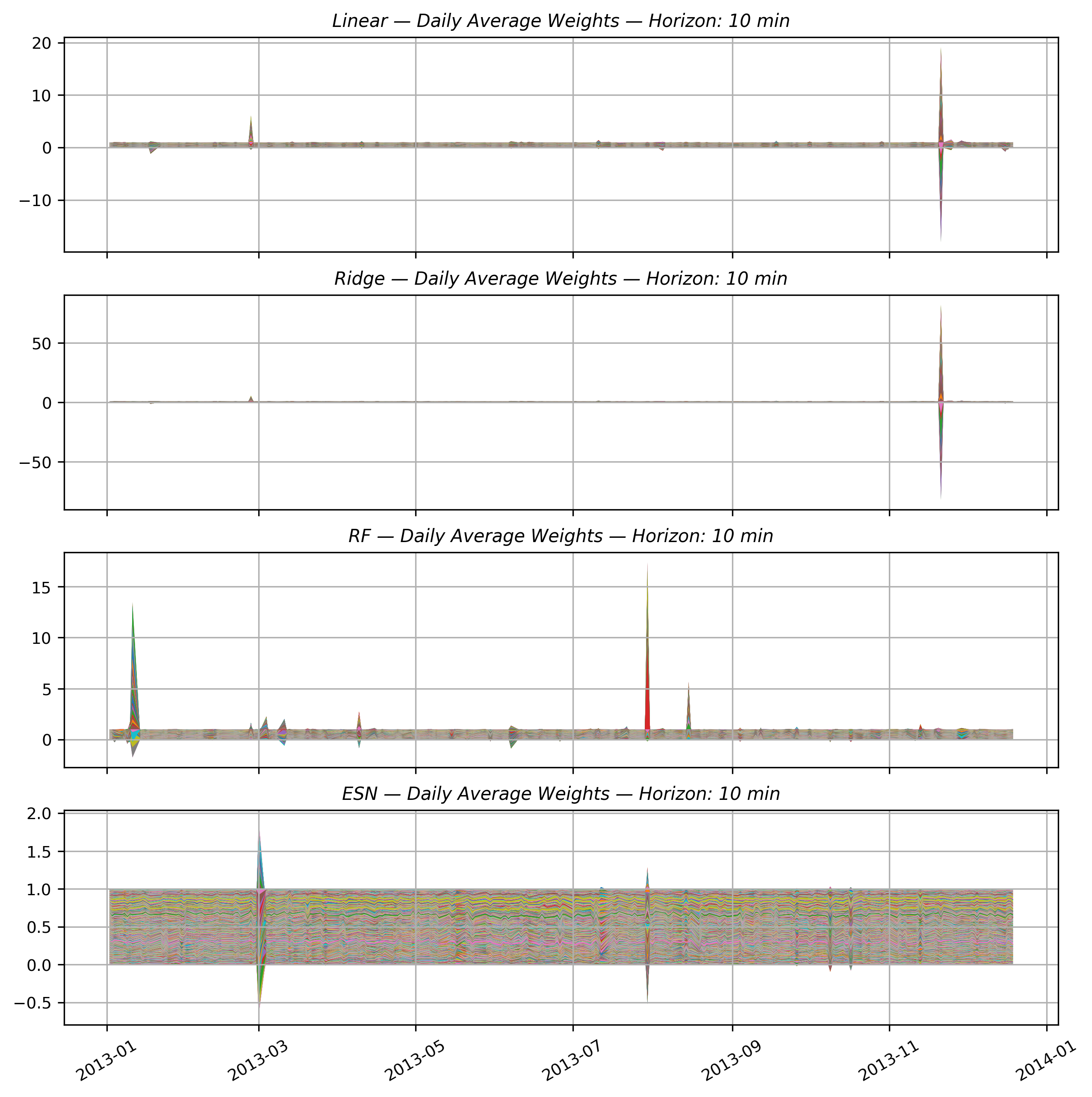}
    \label{fig:daily_average_weights_hr_10m}
\end{figure}

\begin{figure}[p]
    \centering
    \caption{Cumulative Portfolio Returns}
    \begin{subfigure}{0.49\textwidth}
        \includegraphics[width=\textwidth]{figures_portfolio/cumulative_returns_hr_10m.pdf}
    \end{subfigure}
    \hfill
    \begin{subfigure}{0.47\textwidth}
        \includegraphics[width=\textwidth]{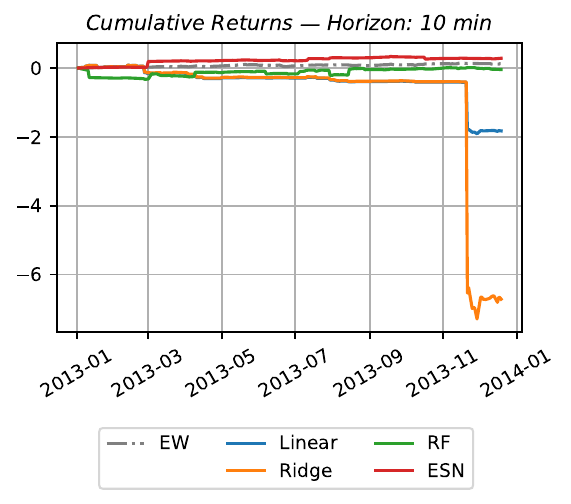}
    \end{subfigure}
    \vspace{5pt}
    \caption*{\small
    \textit{Notes:} Portfolio weights derived with linear, ridge, random forest, and ESN forecasts are computed on identical daily covariance matrix estimates with Ledoit-Wolf nonlinear shrinkage. Cumulative returns are net of transaction costs. Left panel uses a reduced $y$-axis, while the right panel is unrestricted. 
    }
    \label{fig:portfolio_cumret_full}
\end{figure}

\begin{figure}[p]
    \centering
    \caption{Net Cumulative Portfolio Returns}
    \begin{subfigure}{0.49\textwidth}
        \includegraphics[width=\textwidth]{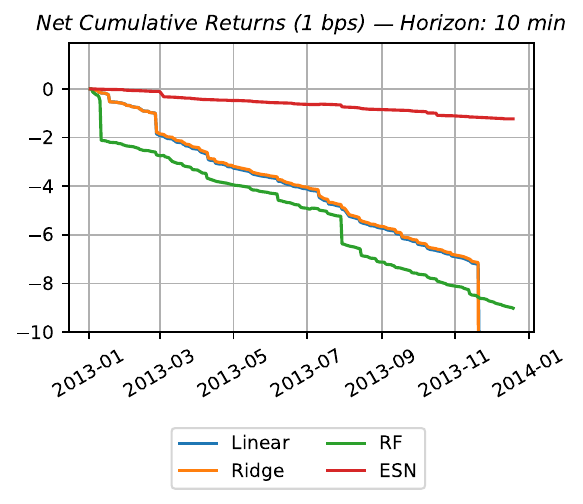}
    \end{subfigure}
    \hfill
    \begin{subfigure}{0.47\textwidth}
        \includegraphics[width=\textwidth]{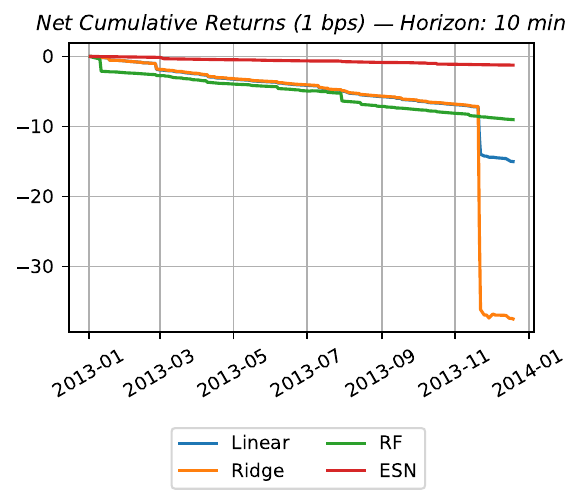}
    \end{subfigure}
     \vspace{5pt}
    \caption*{\small
    \textit{Notes:} Net returns for all portfolios computed with 1bps transaction cost from cumulative returns and turnovers, cf. Figure~\ref{fig:portfolio_cumret_turnover}. Left panel uses a reduced $y$-axis, while the right panel is unrestricted.
    }
    \label{fig:portfolio_net_cumreturns}
\end{figure}

\end{document}